\documentclass[twocolumn]{IEEEtran}
\usepackage{amsmath}
\usepackage{latexsym}
\usepackage{amssymb}
\usepackage{bm}
\usepackage[dvips]{graphicx}

\newtheorem{thm}    {Theorem}
\newtheorem{lem}     {Lemma}
\newtheorem{cor}  {Corollary}
\newtheorem{prop}        {Proposition}
\newtheorem{rem}     {Remark}

\newcommand{\defeq}{\stackrel{\rm def}{=}}
\def\real{\mathbb R}
\def\vlimsup{\varlimsup}
\def\vliminf{\varliminf}
\def\Label#1{\label{#1}\ [\ #1\ ]\ }
\def\Label{\label}
\def\rE{{\rm E}}
\def\cM{{\cal M}}
\begin{document}
\title{Second order asymptotics in\\ fixed-length source coding and
intrinsic randomness}
\author{
Masahito Hayashi
\thanks{
M. Hayashi is with Quantum Computation and Information Project, ERATO, JST,
5-28-3, Hongo, Bunkyo-ku, Tokyo, 113-0033, Japan.
(e-mail: masahito@qci.jst.go.jp)
}}
\date{}
\maketitle

\begin{abstract}
Second order asymptotics of
fixed-length source coding and 
intrinsic randomness is discussed with a constant error constraint.
There was a difference between optimal rates of fixed-length source coding and 
intrinsic randomness, which never occurred in 
the first order asymptotics.
In addition, the relation between uniform distribution and 
compressed data is discussed based on this fact.
These results are valid for general information sources as well as
independent and identical distributions.
A universal code attaining the second order optimal rate 
is also constructed.
\end{abstract}

\section{Introduction}
\PARstart{M}{any} researchers believe that 
any sufficiently compressed data approaches a 
uniform random number.
This conjecture is called Folklore for source coding
(Han \cite{Han2}).
The main reason for this conjecture seems to be
the fact that the optimal limits of both rates
coincide with the entropy rate: that is, 
the optimal compression length equals
the optimal length of intrinsic randomness 
(uniform random number generation)
in the asymptotic first order.
There is, however, no research comparing 
them in the asymptotic second order
even though some researchers treat
the second order asymptotics for variable-length
source coding \cite{Kont,CB}.
In this paper, taking account of the asymptotic second order,
we compare them in the case of  the general information source in the fixed-length
setting. 
Especially, 
we show by application to the case of the independent and identical distribution
(i.i.d.), that
the size of compression is greater than
the one of intrinsic randomness
with respect to the asymptotic second order.
This fact implies that data generated by the fixed-length source coding
is not a uniform random number.

Details of the above discussion are as follows.
The size of generated data is one 
of the main points in data compression and
intrinsic randomness.
In the asymptotic setting, by approximating the size $M_n$ as
$M_n\cong e^{na}$,
we usually focus on the exponential component (exponent) $a$.
Smaller size is better in data compression,
but larger size is better in intrinsic randomness.
Both optimal exponents $a$ coincide.
However, as will be shown in this paper,
the size $M_n$ can be approximated
as $M_n\cong e^{na +\sqrt{n}b}$.
In this paper, we call the issue concerning the coefficient $a$ of $n$
the first order asymptotics,
and the issue concerning the coefficient $b$ of $\sqrt{n}$
the second order asymptotics.
When the information source is the independent and identical distribution
$P^n$ of a probability distribution $P$,
the optimal first coefficient is the entropy $H(P)$ in both settings.
In this paper,
we treat the optimization of the second coefficient $b$ 
for general information sources.
In particular, we treat intrinsic randomness by using half of 
the variational distance. 
These two coefficients do not coincide with each other in many cases. 
In particular, 
these optimal second order coefficients depend on the allowable error
even in the i.i.d. case.
(Conversely, it is known that
these optimal first order coefficients are independent of the allowable error
in the i.i.d.\ case when the allowable error is constant.) 
If the allowable error is less than $1/2$,
the optimal second order coefficient for source coding
is strictly larger than the one for intrinsic randomness.
As a consequence, when the constraint error for source coding
is sufficiently small,
the compressed random number is different from the uniform random number.
Hence, there exists a trade-off relation between
the error of compression and the error of intrinsic randomness. 

However, Han \cite{Han2,Han1,Han3} showed that
the compressed data achieving the optimal rate 
is `almost' uniform random at least in the i.i.d.\ case
in the fixed-length compression.
Visweswariah {\it et al.}\cite{VKV}
and Han \& Uchida \cite{HU} also treated a similar problem
in the variable-length setting.
One may think that Han's result contradicts our result,
but there is no contradiction.
This is because Han's error criterion
between the obtained distribution and the true uniform distribution 
is based on normalized KL-divergence (\ref{eq68}), 
and is not as restrictive as our criterion.
Thus, the distribution of the compressed data may not be different from the 
uniform distribution under our criterion
even if it is `almost' the uniform distribution under his criterion.
Indeed, Han \cite{Han3} has already stated in his conclusion
that if we adopt the variational distance,
the compressed data is different from the uniform random number
in the case of the stationary ergodic source.
However, in this paper, using the results of second order asymptotics,
we succeeded in deriving the tight trade-off relation
between the variational distance from the uniform distribution
and decoding error probability of the fixed-length compression
in the asymptotic setting.
Further, when we adopt KL-divergence divided by $\sqrt{n}$
instead of normalized KL-divergence,
the compressed data is different from the uniform random number.
Hence, the speed of convergence of normalized KL-divergence to $0$ 
is essential.

In this paper, we use the information spectrum method
mainly formulated by Han\cite{Han1}.
We treat the general information source, which is the
general sequence $\{p_n\}$ of probability distributions without
structure.
This method enables us to characterize the asymptotic performance only
with the random variable $\frac{1}{n}\log p_n$ (the logarithm of likelihood)
without any further assumption.
In order to treat the i.i.d.\ case based on the above general result,
it is sufficient to calculate the asymptotic behavior 
of the random variable $\frac{1}{n}\log p_n$.
Moreover, the information spectrum method leads us to 
treat the second order asymptotics 
in a manner parallel to the first order asymptotics, whose large part
is known.
That is, if we can suitably formulate theorems in the 
second order asymptotics
and establish an appropriate relation between 
the first order asymptotics and the second order asymptotics,
we can easily extend proofs concerning the first order asymptotics
to those of the second order asymptotics.
This is because the technique used in the information spectrum method
is quite universal.
Thus, the discussion of the first order asymptotics
plays an important role in our proof of some important theorems in 
the second order asymptotics.
Therefore, we give proofs of some theorems in
the first order asymptotics even though they are known.
This treatment produces 
short proofs of main theorems for the second order asymptotics
with reference to the corresponding proofs on the 
first order asymptotics.

While we referred the i.i.d. case in the above discussion, 
the Markovian case also has a similar asymptotic structure.
That is, the limiting distribution of the logarithm of likelihood 
is equal to normal distribution.
Hence, we have the same conclusion concerning Folklore for source coding
in the Markovian case.
Moreover, we construct a fixed-length source code
that attains the optimal rate 
up to the second order asymptotics,
{\em i.e.}, a universal fixed-length source code.
We also prove the existence of a similar universal operation
for intrinsic randomness.
Further, in Section VI-A, 
we derived the optimal generation rate of intrinsic randomness
under the constant constraint concerning the normalized KL-divergence,
which was mentioned as an open problem in Han's textbook\cite{Han1}.

Finally, we should remark that the second order asymptotics 
correspond to the central limit theorem in the i.i.d.\ case
while the first order asymptotics corresponds to the
law of large numbers.
But, in statistics,
the first order asymptotics corresponds to
the central limit theorem.
Concerning variable-length source coding,
the second order asymptotics 
corresponds to the central limit theorem,
but its order is $\log n$.
As seen in sections VIII and \ref{22-1},
the application of this theorem to 
variable- and fixed-length source coding
is different.

This paper is organized as follows.
We explain some notations for
the information spectrum method in the first 
and the second order asymptotics in section II.
We treat the first order asymptotics
of fixed-length source coding and 
intrinsic randomness based on variational distance in section III,
some of which are known.
For the comparisons with several preceding results,
we treat several versions of the optimal rate in this section.
The second order asymptotics in both settings
are discussed as the main result in section IV.
We discuss the relation between 
the second order asymptotics 
and Folklore for source coding based on variational distance in section V.
In addition, we discuss intrinsic randomness based on KL-divergence,
and the relation between Han\cite{Han1}'s criterion 
and the second order asymptotics in section VI.
For comparison with Han\cite{Han1}'s result, we 
treat intrinsic randomness under another KL-divergence criterion, 
in which the input distributions of KL-divergence are exchanged.
In section VII, the Markovian case is discussed.
A universal fixed-length source code and 
a universal operation for intrinsic randomness
are treated in section VIII.
All proofs are given in section IX.
\section{Notations of information spectrum}
In this paper, we treat general information source.
Through this treatment, we can understand the essential 
properties of problems discussed in this paper.
First, we focus on a sequence of probability spaces
$\{\Omega_n\}_{n=1}^\infty$ and a sequence of probability 
distributions $\overline{p}\defeq
\{p_n\}_{n=1}^\infty$ on them.
The asymptotic behavior of the the logarithm of likelihood 
can be characterized by the following known quantities 
\begin{align*}
\underline{H}(\epsilon|\overline{p})
&\defeq \inf_a 
\{a|
\vlimsup p_n \{ - \frac{1}{n}\log p_n (\omega) < a \} 
\ge \epsilon \}\\
&= \sup_a 
\{a|
\vlimsup p_n \{ - \frac{1}{n}\log p_n (\omega) <  a \} 
< \epsilon \},\\
\overline{H}(\epsilon|\overline{p})
&\defeq 
\inf_a \{a|
\vliminf p_n \{ - \frac{1}{n}\log p_n (\omega) < a \} 
\ge \epsilon \}\\
&=\sup_a \{a|
\vliminf p_n \{ - \frac{1}{n}\log p_n (\omega) < a \} 
< \epsilon \},
\end{align*}
for $0 < \epsilon \le 1$, and 
\begin{align*}
\underline{H}_+(\epsilon|\overline{p})
&\defeq \inf_a 
\{a|
\vlimsup p_n \{ - \frac{1}{n}\log p_n (\omega) < a \} 
> \epsilon \}\\
&= \sup_a 
\{a|
\vlimsup p_n \{ - \frac{1}{n}\log p_n (\omega) <  a \} 
\le \epsilon \},\\
\overline{H}_+(\epsilon|\overline{p})
&\defeq 
\inf_a \{a|
\vliminf p_n \{ - \frac{1}{n}\log p_n (\omega) < a \} 
> \epsilon \}\\
&=\sup_a \{a|
\vliminf p_n \{ - \frac{1}{n}\log p_n (\omega) < a \} 
\le \epsilon \},
\end{align*}
for $0 \le \epsilon < 1$, 
where the $\omega$ is an element of the probability
space $\Omega_n$.

For example, 
when the probability $p_n$ is the $n$-th independent and identical
distribution (i.i.d.) $P^n$ of $P$,
the law of large numbers guarantees that
these quantities coincide with entropy $H(P)
\defeq - \sum_\omega P(\omega) \log P(\omega)$.
Therefore, for a more detailed description of 
asymptotic behavior,
we introduce the following quantities.
\begin{align*}
 \underline{H}(\epsilon,a|\overline{p})
\defeq& \inf_b
\{b|
\vlimsup p_n \{ - \frac{1}{n}\log p_n (\omega_n) < a+ \frac{b}{\sqrt{n}}  
\} \ge \epsilon \}\\
= &\sup_b
\{b|
\vlimsup p_n \{ - \frac{1}{n}\log p_n (\omega_n) <  a + \frac{b}{\sqrt{n}} \} 
< \epsilon \} ,\\
\overline{H}(\epsilon,a|\overline{p})
\defeq &
\inf_b \{b|
\vliminf p_n \{ - \frac{1}{n}\log p_n (\omega_n) < a + \frac{b}{\sqrt{n}} \} 
\ge \epsilon \}\\
=&\sup_b \{b|
\vliminf p_n \{ - \frac{1}{n}\log p_n (\omega_n) < a + \frac{b}{\sqrt{n}} \} 
< \epsilon \},
\end{align*}
for $0 < \epsilon \le 1$.
Similarly, 
$\underline{H}_+(\epsilon,a|\overline{p})$
and 
$\overline{H}_+(\epsilon,a|\overline{p})$
are defined for
$0 \le \epsilon < 1$.
When the distribution $p_n$ is the i.i.d.\ $P^n$ of $P$,
the central limit theorem guarantees that
$ \sqrt{n}(- \frac{1}{n}\log P^n (\omega_n) - H(P))$
obeys the 
normal distribution with expectation $0$ and  variance 
$V_P= \sum_\omega P(\omega) ( -\log P(\omega) - H(P) )^2$.
Therefore, 
by using the distribution function $\Phi$ for
the standard normal distribution
(with expectation $0$ and the variance $1$):
\begin{align*}
\Phi(x)\defeq \int_{-\infty}^{x}
\frac{1}{\sqrt{2\pi}}
e^{- x^2/2}\,d x,
\end{align*}
we can express the above quantities as follows:
\begin{align}
&\underline{H}(\epsilon,H(P) |\overline{P})
=\overline{H}(\epsilon,H(P) |\overline{P})\nonumber \\
=&\underline{H}_+(\epsilon,H(P) |\overline{P})
=\overline{H}_+(\epsilon,H(P) |\overline{P})
= \sqrt{V_P}\Phi^{-1}(\epsilon) \Label{25},
\end{align}
where
$\overline{P}= \{P^n\}$.

In the following, we discuss 
the relation between the above mentioned quantities,
fixed-length source coding, and intrinsic randomness. 

\begin{figure}[htbp]
\begin{center}
\scalebox{1.0}{\includegraphics[scale=0.35]{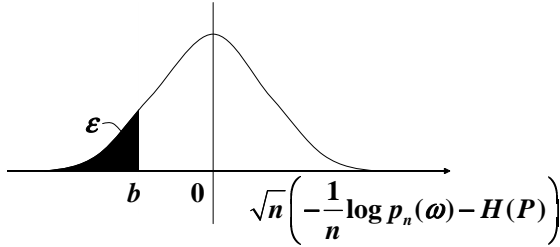}}
\end{center}
\caption{The limiting distribution of 
the logarithm of likelihood}
\label{graph1}
\end{figure}%

\section{First order asymptotics}
\subsection{Fixed-length source coding}
In fixed-length source coding,
first we fix a set of integers
${\cal M}_n\defeq \{ 1, \ldots, M_n\}$.
The transformation from the output $\omega \in \Omega_n$ 
to an element of the set ${\cal M}_n$ is
described by a map $\phi_n:\Omega_n  \to {\cal M}_n$,
which is called {\em encoding}.

\begin{figure}[htbp]
\begin{center}
\scalebox{1.0}{\includegraphics[scale=0.35]{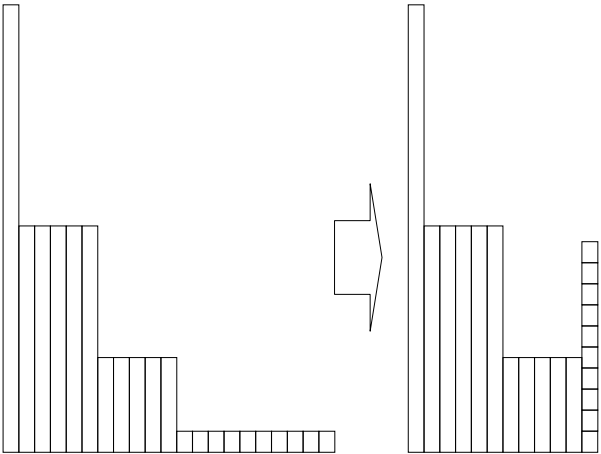}}
\end{center}
\caption{Encoding operation in source coding}
\label{graph2}
\end{figure}%

The operation recovering the original output $\omega$
from the element of ${\cal M}_n$ is 
described by a map $\psi_n : {\cal M}_n\to \Omega_n$,
and is called {\em decoding}.
We call the triple $\Phi_n({\cal M}_n,\phi_n,\psi_n)$ a {\em code}, and 
evaluate its performance by its size
$|\Phi_n|\defeq |{\cal M}_n|= M_n$ and error probability:
\begin{align*}
\varepsilon_{p_n}(\Phi_n) \defeq 
p_n
\{
\omega\in \Omega_n| \psi_n \circ\phi_n (\omega)\neq \omega
\}.
\end{align*} 
When we do not need to express the distribution of information source
$p_n$,
we simplify $\varepsilon_{p_n}(\Phi_n)$ to
$\varepsilon(\Phi_n)$.
In order to discuss
the asymptotic bound of compression rate
under the constant constraint on the
error probability,
we focus on the following values:
\begin{align*}
R(\epsilon|\overline{p})& \defeq
\inf_{\{\Phi_n\}}
\{
\vlimsup \frac{1}{n} \log |\Phi_n| |
\vlimsup \varepsilon(\Phi_n) \le \epsilon\},\\
R^\dagger(\epsilon|\overline{p})& \defeq
\inf_{\{\Phi_n\}}
\{
\vlimsup \frac{1}{n} \log |\Phi_n| |
\vliminf \varepsilon(\Phi_n) \le \epsilon\},\\
R^\ddagger(\epsilon|\overline{p})& \defeq
\inf_{\{\Phi_n\}}
\{
\vliminf \frac{1}{n} \log |\Phi_n| |
\vlimsup \varepsilon(\Phi_n) \le \epsilon\},
\end{align*}
for $0 \le \epsilon <1$, and
\begin{align*}
R_+(\epsilon|\overline{p})& \defeq
\inf_{\{\Phi_n\}}
\{
\vlimsup \frac{1}{n} \log |\Phi_n| |
\vlimsup \varepsilon(\Phi_n) < \epsilon\},\\
R_+^\dagger(\epsilon|\overline{p})& \defeq
\inf_{\{\Phi_n\}}
\{
\vlimsup \frac{1}{n} \log |\Phi_n| |
\vliminf \varepsilon(\Phi_n) < \epsilon\},\\
R_+^\ddagger(\epsilon|\overline{p})& \defeq
\inf_{\{\Phi_n\}}
\{
\vliminf \frac{1}{n} \log |\Phi_n| |
\vlimsup \varepsilon(\Phi_n) < \epsilon\},
\end{align*}
for $0 < \epsilon \le 1$.
Further, 
as intermediate quantities, we define
\begin{align*}
\tilde{R}(\epsilon|\overline{p})
& \defeq
\inf_{\{\Phi_n\}}
\{
\vlimsup \frac{1}{n} \log |\Phi_n| |
\varepsilon(\Phi_n) \le \epsilon,~ \forall n\}\\
& =
\inf_{\{\Phi_n\}}
\{
\vlimsup \frac{1}{n} \log |\Phi_n| |
\exists N 
\varepsilon(\Phi_n) \le \epsilon,~ \forall n \ge N\},\\
\tilde{R}^\dagger(\epsilon|\overline{p})& \defeq
\inf_{\{\Phi_n\}}
\left\{
\vlimsup \frac{1}{n} \log |\Phi_n| 
\left|
\begin{array}{l}
\varepsilon(\Phi_n) \le \epsilon \\
\hbox{for infinitely many }n
\end{array}
\right.\right\},\\
\tilde{R}^\ddagger(\epsilon|\overline{p})
& \defeq
\inf_{\{\Phi_n\}}
\{
\vliminf \frac{1}{n} \log |\Phi_n| |
\varepsilon(\Phi_n) \le \epsilon\} \\
& =
\inf_{\{\Phi_n\}}
\{
\vliminf \frac{1}{n} \log |\Phi_n| |\exists N
\varepsilon(\Phi_n) \le \epsilon,~ \forall n \ge N\},
\end{align*}
for $0 < \epsilon < 1$.
Here, in order to see the relation with existing results,
we defined many versions of the optimal coding length.
The following relations follow from their definitions:
\begin{align}
R(\epsilon|\overline{p})
&\le \tilde{R}(\epsilon|\overline{p})
\le R_+(\epsilon|\overline{p}),\label{8-5-1-a}
\\
R^\dagger(\epsilon|\overline{p})
& \le \tilde{R}^\dagger(\epsilon|\overline{p})
\le R_+^\dagger(\epsilon|\overline{p}),\label{8-5-2-a}\\
R^\ddagger(\epsilon|\overline{p})
& \le \tilde{R}^\ddagger(\epsilon|\overline{p})
\le R_+^\ddagger(\epsilon|\overline{p}),\label{8-5-3}
\end{align}
for $0 < \epsilon <1$.

Concerning these quantities, 
the following theorem holds.
\begin{thm}\Label{th1}
Han\cite[Theorem 1.6.1]{Han1},
Steinberg \& Verd\'{u}\cite{S-V},
Chen \& Alajaji \cite{CA},
Nagaoka \& Hayashi \cite{N-H}
The relations
\begin{align}
&R(1-\epsilon|\overline{p})
= \overline{H}(\epsilon|\overline{p}), \label{8-5-20}\\
&R^\dagger(1-\epsilon|\overline{p})
=R^\ddagger(1-\epsilon|\overline{p})
= \underline{H}(\epsilon|\overline{p}) \label{8-5-21}
\end{align}
hold for $0 \le \epsilon < 1$, and the relations 
\begin{align}
&R_+(1-\epsilon|\overline{p})
= \overline{H}_+(\epsilon|\overline{p}), \label{8-5-22}\\
& R_+^\dagger(1-\epsilon|\overline{p})
=R_+^\ddagger(1-\epsilon|\overline{p})
= \underline{H}_+(\epsilon|\overline{p})\label{8-5-23}
\end{align}
hold for $0 < \epsilon \le 1$.
\end{thm}

By using the relations (\ref{8-5-1-a}), (\ref{8-5-2-a}), and (\ref{8-5-3}),
$\tilde{R}(\epsilon|\overline{p})$, 
$\tilde{R}^\dagger(\epsilon|\overline{p})$, and
$\tilde{R}^\ddagger(\epsilon|\overline{p})$
are characterized as follows.
\begin{cor}
\begin{align}
&\overline{H}(\epsilon|\overline{p}) 
\le\tilde{R}(1-\epsilon|\overline{p})
\le \overline{H}_+(\epsilon|\overline{p}), \label{8-5-4}\\
&\underline{H}(\epsilon|\overline{p})
\le \tilde{R}^\dagger(1-\epsilon|\overline{p})
\le \underline{H}_+(\epsilon|\overline{p}),\label{8-5-5-a}\\
&\underline{H}(\epsilon|\overline{p})
\le \tilde{R}^\ddagger(1-\epsilon|\overline{p})
\le \underline{H}_+(\epsilon|\overline{p}).\label{8-5-6}
\end{align}
\end{cor}

\begin{rem}\rm
Historically,
Steinberg \& Verd\'{u}\cite{S-V}
derived (\ref{8-5-4}), and
Chen \& Alajaji \cite{CA} did (\ref{8-5-5-a}).
Han \cite{Han1} proved the equation
$R(1-\epsilon|\overline{p})= \overline{H}(\epsilon|\overline{p})$.
Following these results, Nagaoka \& Hayashi \cite{N-H} proved 
$R_+^\dagger(1-\epsilon|\overline{p})=\underline{H}_+(\epsilon|\overline{p})$.
Other relations are proved for the first time in this paper.
\end{rem}

The bounds $R_+^\dagger(1|\overline{p})$ and 
$R_+^\ddagger(1|\overline{p})$
are shortest among the above bounds
because 
$R(\epsilon|\overline{p})$, 
$R^\dagger(\epsilon|\overline{p})$, 
$R^\ddagger(\epsilon|\overline{p})$,
$\tilde{R}(\epsilon|\overline{p})$, 
$\tilde{R}^\dagger(\epsilon|\overline{p})$, and
$\tilde{R}^\ddagger(\epsilon|\overline{p})$
are not defined for $epsilon =1$.
Hence, 
the bounds $R_+^\dagger(1|\overline{p})$ and 
$R_+^\ddagger(1|\overline{p})$
are used in the discussion concerning strong converse property.

\subsection{Intrinsic randomness}
Next, we consider the problem of constructing approximately
the uniform probability distribution from a biased 
probability distribution $p_n$ on $\Omega_n$.
We call this problem intrinsic randomness,
and discuss it based on (half) the variational distance in this
section.
Our operation is described by
the pair of size $M_n$ of the target uniform probability distribution
and the map $\phi_n$ from $\Omega_n$ to ${\cal M}_n \defeq\{1, \ldots, M_n\}$.

\begin{figure}[htbp]
\begin{center}
\scalebox{1.0}{\includegraphics[scale=0.35]{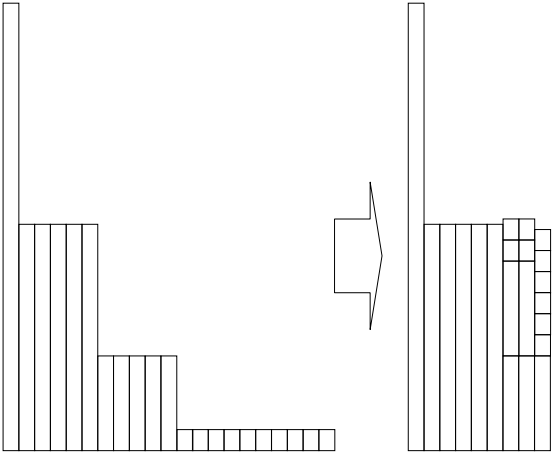}}
\end{center}
\caption{Typical 
operation of intrinsic randomness}
\label{graph3}
\end{figure}%

Performance of $\Psi_n=({\cal M}_n,\phi_n)$ 
is characterized by
the size $|\Psi_n|\defeq M_n$ and a half of the variational distance between the target distribution 
and the constructed distribution:
\begin{align}
\varepsilon_{p_n}(\Psi_n)\defeq
d(p_n \circ \phi_n^{-1}, p_{U,{\cal M}_n}) ,
\Label{20-1}
\end{align}
where
$d(p,q)\defeq \frac{1}{2}\sum_\omega |p(\omega)-q(\omega)|$
and 
$p_{U,{\cal S}}$ is the uniform distribution on ${\cal S}$.
When we do not need to express the distribution of information source,
$p_n$,
we simplify $\varepsilon_{p_n}(\Psi_n)$ to
$\varepsilon(\Psi_n)$.
Under the condition that this distance is less than $\epsilon$
the optimal size is asymptotically characterized as follows:
\begin{align*}
S(\epsilon|\overline{p})& \defeq
\sup_{\{\Psi_n\}}
\{
\vliminf \frac{1}{n} \log |\Psi_n| |
\vlimsup \varepsilon(\Psi_n) \,< \epsilon\},\\
S^\dagger(\epsilon|\overline{p})& \defeq
\sup_{\{\Psi_n\}}
\{
\vliminf \frac{1}{n} \log |\Psi_n| |
\vliminf \varepsilon(\Psi_n) \,< \epsilon\},\\
S^{\ddagger}(\epsilon|\overline{p})& \defeq
\sup_{\{\Psi_n\}}
\{
\vlimsup \frac{1}{n} \log |\Psi_n| |
\vlimsup \varepsilon(\Psi_n) \,< \epsilon\},
\end{align*}
for $0 < \epsilon \le 1$, and 
\begin{align*}
S_+(\epsilon|\overline{p})& \defeq
\sup_{\{\Psi_n\}}
\{
\vliminf \frac{1}{n} \log |\Psi_n| |
\vlimsup \varepsilon(\Psi_n) \le \epsilon\},\\
S_+^\dagger(\epsilon|\overline{p})& \defeq
\sup_{\{\Psi_n\}}
\{
\vliminf \frac{1}{n} \log |\Psi_n| |
\vliminf \varepsilon(\Psi_n) \le \epsilon\},\\
S_+^{\ddagger}(\epsilon|\overline{p})& \defeq
\sup_{\{\Psi_n\}}
\{
\vlimsup \frac{1}{n} \log |\Psi_n| |
\vlimsup \varepsilon(\Psi_n) \le \epsilon\},
\end{align*}
for $0 \le \epsilon < 1$.
As intermediate quantities, 
\begin{align*}
\tilde{S}(\epsilon|\overline{p})& \defeq
\sup_{\{\Psi_n\}}
\{
\vliminf \frac{1}{n} \log |\Psi_n| |
\varepsilon(\Psi_n) \le \epsilon\},\\
\tilde{S}^\dagger(\epsilon|\overline{p})& \defeq
\sup_{\{\Psi_n\}}
\left\{
\vliminf \frac{1}{n} \log |\Psi_n| \left|
\begin{array}{l}
\varepsilon(\Psi_n) \le \epsilon\\
\hbox{for infinitely many } n
\end{array}
\right.\right\},\\
\tilde{S}^{\ddagger}(\epsilon|\overline{p})& \defeq
\sup_{\{\Psi_n\}}
\{
\vlimsup \frac{1}{n} \log |\Psi_n| |
\varepsilon(\Psi_n) \le \epsilon\}
\end{align*}
are defined for 
$0 < \epsilon < 1$.
Similarly, we obtain the following trivial relations:
\begin{align}
S(\epsilon|\overline{p})
&\le \tilde{S}(\epsilon|\overline{p})
\le S_+(\epsilon|\overline{p}),\label{8-5-10}
\\
S^\dagger(\epsilon|\overline{p})
& \le \tilde{S}^\dagger(\epsilon|\overline{p})
\le S_+^\dagger(\epsilon|\overline{p}),\label{8-5-11}\\
S^\ddagger(\epsilon|\overline{p})
& \le \tilde{S}^\ddagger(\epsilon|\overline{p})
\le S_+^\ddagger(\epsilon|\overline{p}),\label{8-5-12}
\end{align}
for $0 < \epsilon <1$.

These quantities are characterized by the following 
theorem.
\begin{thm}\Label{th3}Han\cite[Theorem 2.4.2]{Han1}
The relations
\begin{align}
S(\epsilon|\overline{p})
= \underline{H}(\epsilon|\overline{p}), \quad
S^\dagger(\epsilon|\overline{p})
=S^\ddagger(\epsilon|\overline{p})
= \overline{H}(\epsilon|\overline{p})\label{8-5-15}
\end{align}
hold for $0 < \epsilon \le 1$,
and the relations
\begin{align}
S_+(\epsilon|\overline{p})
= \underline{H}_+(\epsilon|\overline{p}), \quad
S_+^\dagger(\epsilon|\overline{p})
=S_+^\ddagger(\epsilon|\overline{p})
= \overline{H}_+(\epsilon|\overline{p})\label{8-5-16}
\end{align}
hold for $0 \le \epsilon < 1$.
\end{thm}
Similarly, the following corollary holds.
\begin{cor}
The relations 
\begin{align}
&\underline{H}(\epsilon|\overline{p})
\le \tilde{S}(\epsilon|\overline{p})
\le \underline{H}_+(\epsilon|\overline{p}), \\
&\overline{H}(\epsilon|\overline{p})
\le \tilde{S}^\dagger(\epsilon|\overline{p})
\le \overline{H}_+(\epsilon|\overline{p}) ,\\
&\overline{H}(\epsilon|\overline{p})
\le \tilde{S}^\ddagger(\epsilon|\overline{p})
\le \overline{H}_+(\epsilon|\overline{p})
\end{align}
hold for $0 < \epsilon < 1$.
\end{cor}

\begin{rem}\rm
Han\cite{Han1} proved only the first equation of (\ref{8-5-15}).
Other equations are proved for the first time in this paper.
\end{rem}

In the following,
in order to treat Folklore for source coding,
we focus on the operation $\Psi_n=({\cal M}_n, \phi_n)$
defined from the code 
$\Phi_n=({\cal M}_n, \phi_n,\psi_n)$.
For fixed real numbers $\epsilon$ and $\epsilon'$ satisfying
$0 \le \epsilon, \epsilon' \,< 1$,
we consider whether 
there exist
codes $\Phi_n=({\cal M}_n, \phi_n,\psi_n)$ such that
\begin{align}
\vlimsup \varepsilon (\Phi_n) \le \epsilon, \quad
\vlimsup \varepsilon (\Psi_n) \le \epsilon' \Label{eq20}.
\end{align}
If there exists a sequence of codes $\{\Phi_n\}$ satisfying 
the above conditions,
the inequalities
\begin{align*}
&\underline{H}(\epsilon'|\overline{p})
= S(\epsilon'| \overline{p})
\ge \vliminf \frac{1}{n}\log M_n 
\ge R(\epsilon| \overline{p})
=\underline{H}(1-\epsilon|\overline{p}),\\
& \overline{H}(\epsilon'|\overline{p})
= S^\ddagger(\epsilon'| \overline{p})
\ge 
\vlimsup \frac{1}{n}\log M_n
\ge R^\ddagger(\epsilon| \overline{p})
=\overline{H}(1-\epsilon|\overline{p})
\end{align*}
hold.
Therefore, we obtain the following necessary 
condition for the existence of 
$\{\Phi_n\}$ satisfying (\ref{eq20}):
\begin{align}
\underline{H}(\epsilon'|\overline{p})\ge 
\underline{H}(1-\epsilon|\overline{p}), \quad
\overline{H}(\epsilon'|\overline{p})\ge 
\overline{H}(1-\epsilon|\overline{p})\Label{22}.
\end{align}
Thus, the necessary condition (\ref{22}) is satisfied 
in the case of i.i.d. $P^n$
because these quantities coincide with the entropy $H(P)$.

However, the above discussion is not sufficient,
because,
as is shown based on the second order asymptotics,
a stronger necessary condition exists.
\section{Second order asymptotics}
Next, we proceed to the second order asymptotics,
which is very useful for obtaining the stronger necessary condition
than (\ref{22}).
Since
these values
$\underline{H}(\epsilon'|\overline{p}),\overline{H}(\epsilon|\overline{p})$
are independent of $\epsilon$ in the i.i.d.\ case,
we introduce the following values
for treatment of the dependence of $\epsilon$:
\begin{align*}
R(\epsilon, a|\overline{p})
 \defeq&
\inf_{\{\Phi_n\}}
\{
\vlimsup
\frac{1}{\sqrt{n}} \log \frac{|\Phi_n|}{e^{na}}
|
\vlimsup \varepsilon(\Phi_n) \le \epsilon\},\\
R^\dagger(\epsilon,a|\overline{p})
\defeq&
\inf_{\{\Phi_n\}}
\{
\vlimsup
\frac{1}{\sqrt{n}} \log \frac{|\Phi_n|}{e^{na}}
|
\vliminf \varepsilon(\Phi_n) \le \epsilon\},\\
R^\ddagger(\epsilon, a|\overline{p})
 \defeq&
\inf_{\{\Phi_n\}}
\{
\vliminf
\frac{1}{\sqrt{n}} \log \frac{|\Phi_n|}{e^{na}}
|
\vlimsup \varepsilon(\Phi_n) \le \epsilon\},
\end{align*}
for $0 \le \epsilon < 1$, and 
\begin{align*}
S(\epsilon, a|\overline{p})
 \defeq &
\sup_{\{\Psi_n\}}
\{
\vliminf 
\frac{1}{\sqrt{n}} \log \frac{|\Psi_n|}{e^{na}}
|
\vlimsup \varepsilon(\Psi_n) \le \epsilon\},\\
 S^\dagger(\epsilon,a|\overline{p})
\defeq &
\sup_{\{\Psi_n\}}
\{
\vliminf
\frac{1}{\sqrt{n}} \log \frac{|\Psi_n|}{e^{na}}
|
\vliminf \varepsilon(\Psi_n) \le \epsilon\},\\
S^\ddagger(\epsilon, a|\overline{p})
 \defeq &
\sup_{\{\Psi_n\}}
\{
\vlimsup 
\frac{1}{\sqrt{n}} \log \frac{|\Psi_n|}{e^{na}}
|
\vlimsup \varepsilon(\Psi_n) \le \epsilon\},
\end{align*}
for $0 < \epsilon \le 1$.
While we can define other quantities
$R_+(\epsilon, a|\overline{p})$,
$R_+^\dagger(\epsilon, a|\overline{p})$,
$R_+^\ddagger(\epsilon, a|\overline{p})$,
$S_+(\epsilon, a|\overline{p})$,
$S_+^\ddagger(\epsilon, a|\overline{p})$, and
$S_+^\ddagger(\epsilon, a|\overline{p})$,
we treat only the above values in this section.
This is because the later values can be treated in a similar way.
The following theorem holds.
\begin{thm}\Label{th4}
\begin{align*}
S(\epsilon,a|\overline{p})&= R^\dagger(1-\epsilon,a|\overline{p})
= R^\ddagger(1-\epsilon,a|\overline{p})
= \underline{H}(\epsilon,a|\overline{p}), \\
S^\dagger(\epsilon,a|\overline{p})&= 
S^\ddagger(\epsilon,a|\overline{p})= R(1-\epsilon,a|\overline{p})
= \overline{H}(\epsilon,a|\overline{p}).
\end{align*}
\end{thm}

Especially, in the case of the i.i.d. $P^n$, 
as is characterized in (\ref{25}),
these quantities with $a= H(P)$ depend on $\epsilon$.
\section{Relation to Folklore for source coding}
Next, we apply Theorem \ref{th4} to
the relation between the code 
$\Phi_n=({\cal M}_n, \phi_n,\psi_n)$ 
and the operation $\Psi_n=({\cal M}_n, \phi_n)$. 
When 
\begin{align}
\vlimsup \varepsilon(\Phi_n) = \epsilon, \quad
\vlimsup \varepsilon(\Psi_n) = \epsilon' \Label{eq26},
\end{align}
similar to the previous section, we can 
derive the following inequalities:
\begin{align*}
\underline{H}(\epsilon',a|\overline{p})\ge 
\underline{H}(1-\epsilon,a|\overline{p}), \quad
\overline{H}(\epsilon',a|\overline{p})\ge 
\overline{H}(1-\epsilon,a|\overline{p}).
\end{align*}
Thus, if 
$\overline{H}(\epsilon',a|\overline{p})$ 
or 
$\underline{H}(\epsilon',a|\overline{p})$ 
is continuous with respect to $\epsilon'$ at least
as in the i.i.d.\ case,
the above equation yields 
$\epsilon' \ge 1-\epsilon$.
That is,
the following trade-off holds between 
the error probability of compression and 
the performance of intrinsic randomness:
\begin{align}
\vlimsup \varepsilon(\Phi_n) + \vlimsup\varepsilon(\Psi_n)\ge 1.
\Label{32-}
\end{align}
Therefore, Folklore for source coding does not hold
with respect to variational distance.
In other word, generating completely uniform random numbers 
requires over compression.
Generally, the following theorem holds.
\begin{thm}\Label{thm10}
We define the distance from the uniform distribution as follows:
\begin{align}
\delta(p_n)\defeq 
\min_{{\cal S} \subset \Omega_n}
d(p_n,  p_{U,{\cal S}}).
\end{align}
Then the following inequality holds:
\begin{align}
\varepsilon (\Phi_n) + \varepsilon (\Psi_n) \ge 
\delta(p_n)
\Label{8-4-5}.
\end{align}
\end{thm}
Especially, in the i.i.d.\ case, the quantity $\delta(p_n)$ goes to $1$.
In such a case, the trade-off relation
\begin{align}
\vliminf (\varepsilon (\Phi_n) + \varepsilon (\Psi_n)) \ge 1
\Label{8-4-5-a}
\end{align}
holds.
Furthermore, the above trade-off inequality (\ref{8-4-5-a}) is rigid 
as is indicated by the following theorem.
\begin{thm}\Label{th5}
When the convergence
$\lim_{n \to \infty} 
p_n \{ - \frac{1}{n}\log p_n (\omega_n) < a+ 
\frac{b+\gamma}{\sqrt{n}}  
\}$ is uniform concerning $\gamma$ in an enough small neibourhood of $0$
and 
the relation
\begin{align*}
\lim_{\gamma\to 0}
\lim_{n \to \infty} 
p_n \{ - \frac{1}{n}\log p_n (\omega_n) < a+ 
\frac{b+\gamma}{\sqrt{n}}  
\} = \epsilon
\end{align*}
holds, there exists a sequence of codes $\Phi_n= ({\cal M}_n,\phi_n,\psi_n)$
($\Psi_n= ({\cal M}_n,\phi_n)$)
satisfying the following conditions:
\begin{align}
&\vlimsup \varepsilon(\Phi_n) \le 1- \epsilon,\quad
\vlimsup \varepsilon(\Psi_n) \le \epsilon,\Label{35}\\
&\vliminf \frac{1}{\sqrt{n}}\log \frac{|\Phi_n|}{e^{na}}
=b\Label{36}.
\end{align}
\end{thm}
\section{Intrinsic randomness based on KL-divergence criterion}
\subsection{First order asymptotics}
Next, we discuss intrinsic randomness based on KL-divergence.
Since Han \cite{Han2} discussed 
Folklore for source coding based on KL-divergence criterion,
we need this type of discussion for comparing 
our result and Han's result. 
The first work on intrinsic randomness based on KL-divergence
was done by Vembu \& Verd\'{u}\cite{V-V}.
They focused on the normalized KL-divergence:
\begin{align}
\frac{1}{n}D(p_n\circ \phi_n^{-1}\|p_{U,{\cal M}_n})\Label{eq68},
\end{align}
where $D(p\|q)$ is the KL-divergence $\sum_\omega p(\omega)\log 
\frac{p(\omega)}{q(\omega)}$.
Han \cite{Han2} called the sequence of distributions 
$p_n\circ \phi_n^{-1}$ `almost' uniform random
if the above value goes to $0$.
\begin{prop}\Label{v-v}
Vembu \& Verd\'{u}\cite[Theorem 1]{V-V}
\begin{align}
& S^*(\overline{p})\defeq 
\sup_{\Psi_n}\{
\vliminf \frac{1}{n}\log |\Psi_n|
|
\lim
\frac{1}{n}D(p_n\circ \phi_n^{-1}\|p_{U,{\cal M}_n}) = 0
\}\nonumber
\\
= & \underline{H}(\overline{p}) \defeq
\sup_a 
\{a|
\vlimsup p_n \{ - \frac{1}{n}\log p_n (\omega) <  a \} 
=0 \} .
\end{align}
\end{prop}
In a thorough discussion of the above proposition,
Han \cite{Han2} worked out the following 
proposition concerning Folklore for source coding.
\begin{prop}Han\cite[Theorem 31]{Han2}
The following three conditions for
the sequence $\overline{p}=\{p_n\}$
are equivalent:
\begin{itemize}
\item
When a sequence of codes $\Phi_n=({\cal M}_n,\phi_n,\psi_n)$
satisfies $\varepsilon(\Phi_n) \to 0$,
$\frac{1}{n}\log|{\cal M}_n|
\to \overline{H}(\overline{p})$
then the value (\ref{eq68}) goes to $0$.
\item
There exists a sequence of codes $\Phi_n=({\cal M}_n,\phi_n,\psi_n)$
satisfying $\varepsilon(\Phi_n) \to 0$,
$\frac{1}{n}\log|{\cal M}_n|
\to \overline{H}(\overline{p})$ and
the value (\ref{eq68}) goes to $0$.
\item The sequence $\overline{p}=\{p_n\}$ satisfies 
the strong converse property:
\begin{align}
\underline{H}(\overline{p})= \overline{H}(\overline{p}) \defeq
\inf_a \{a|
\vliminf p_n \{ - \frac{1}{n}\log p_n (\omega) < a \} 
=1 \}.\Label{8-5-1}
\end{align}
\end{itemize}
\end{prop}
In order to discuss Folklore for source coding
in KL-divergence criterion,
we need to generalize Vembu \& Verd\'{u}'s theorem as follows.
\begin{thm}\Label{th9}
Assume that
$\overline{H}(\epsilon|\overline{p})=
\underline{H}(\epsilon|\overline{p})$.
We define 
the probability distribution function $F$ by
\begin{align*}
\int_0^{\underline{H}(\epsilon|\overline{p})}  F(\,d x) = \epsilon.
\end{align*}
Then, the inequality 
\begin{align}
\vliminf\frac{1}{n}D(p_n\circ \phi_n^{-1}\|p_{U,{\cal M}_n})
\ge 
\int_{0}^{a}
(a-x)F(\,d x)\Label{8-5-2-1}
\end{align}
holds, where 
$a=\vliminf \frac{1}{n}\log M_n$.
Furthermore, 
when $\underline{H}(1-\epsilon|\overline{p})= a$,
there exists a sequence of codes $\{\Phi_n\}$ attaining 
the equality of (\ref{8-5-2-1}) and satisfying 
$\lim \varepsilon (\Phi_n)= \epsilon$.
Here, we remark that the inequality (\ref{8-5-2-1})
is equivalent to the inequality:
\begin{align}
\vliminf\frac{1}{n}H(p_n\circ \phi_n^{-1})
\le \int_{0}^{a}
x F(\,d x) + a(1-F(a)) \Label{8-5-5}.
\end{align}
\end{thm}
Note that the following equation follows from the above theorem:
\begin{align*}
& S^*(\delta|\overline{p})\\
\defeq &
\sup_{\{\Psi_n\}}\{
\vliminf \frac{1}{n}\log |\Psi_n|
|
\vlimsup
\frac{1}{n}D(p_n\circ \phi_n^{-1}\|p_{U,{\cal M}_n}) \le \delta
\}
\\
= &\sup_a
\left\{ a \left| \int_0^a (a-x)F(\,d x)\le \delta\right.\right\}.
\end{align*}
\begin{rem}
The characterization $S^*(\delta|\overline{p})$ as a function of $\delta$
was treated as an open problem in Han's textbook \cite{Han1}.
\end{rem}
In the i.i.d. case of probability distribution $P$,
since 
\begin{align*}
\int_0^a (a-x)F(\,d x)=
\left\{
\begin{array}{ll}
a - H(P) & a \ge H(P) \\
0 & a \,< H(P),
\end{array}
\right.
\end{align*}
we obtain
\begin{align*}
S^*(\delta|\overline{P})= H(P)+\delta.
\end{align*}

Next, we focus on the opposite criterion:
\begin{align*}
D(p_{U,{\cal M}_n}\|p_n\circ \phi_n^{-1}),
\end{align*}
and define the following quantities:
\begin{align*}
&S^*_1(\delta|\overline{p}) \\
\defeq &
\sup_{\{\Psi_n\}}\{
\vliminf \frac{1}{n}\log |\Psi_n|
|
\vlimsup
D(p_{U,{\cal M}_n}\|p_n\circ \phi_n^{-1}) \,< \delta
\},\\
& S^*_2(\delta|\overline{p})\\
 \defeq &
\sup_{\{\Psi_n\}}\{
\vliminf \frac{1}{n}\log |\Psi_n|
|
\vlimsup
\frac{1}{n}D(p_{U,{\cal M}_n}\|p_n\circ \phi_n^{-1}) \,< \delta
\}.
\end{align*}
Then, they are characterized as follows:
\begin{thm}\Label{th15}
\begin{align}
S^*_1(\delta|\overline{p})
= \underline{H} (1- e^{-\delta}|\overline{p}).\Label{19-8}
\end{align}
If 
the limit
\begin{align*}
\sigma(a)\defeq \lim \frac{-1}{n}\log p_n \{
\frac{-1}{n}\log p_n(\omega)
\ge a\}
\end{align*}
converges, the relation
\begin{align}
S^*_2(\delta|\overline{p})
&= \sup_a \{ a- \sigma(a)|\sigma(a) \,< \delta\}\Label{8-19}
\end{align}
holds for $\forall \delta \,>0$.
\end{thm}
\begin{rem}
Indeed, Han \cite{Han1}
proved 
a similar relation concerning the fixed-length source coding 
with the constraint for error exponent:
\begin{align}
& \inf_{\{\Phi_n\}}
\{
\vlimsup \frac{1}{n}\log |\Phi_n|
| 
\vliminf \frac{-1}{n} \log \varepsilon (\Phi_n) \ge r\}\nonumber \\
= &
\sup_a \{ a- \underline{\sigma}(a)|\underline{\sigma}(a) \,< r\},
\Label{19-11}
\end{align}
where
\begin{align*}
\underline{\sigma}(a)\defeq
\vliminf \frac{-1}{n}\log p_n \{
\frac{-1}{n}\log p_n(\omega)
\ge a\}.
\end{align*}
Moreover, Nagaoka and Hayashi \cite{N-H} proved that
equation (\ref{19-11}) holds when
we define $\underline{\sigma}(a)$ by
\begin{align}
\underline{\sigma}(a)\defeq
\vliminf \frac{-1}{n}\log p_n \{
\frac{-1}{n}\log p_n(\omega)
> a\}. \Label{31-1}
\end{align}
Hence, when the limit $\sigma(a)$ exists,
equation (\ref{8-19}) holds with replacing 
$\sigma(a)$ by (\ref{31-1}).

Further, Hayashi \cite{ha} showed that 
when the limit $\sigma(a)$ exists,
$\sup_a \{ a- \sigma(a)|\sigma(a) \le r\}$
is equal to the bound of generation rate of maximally 
entangled state with the exponential constraint for success probability
with the correspondence of each probability 
to the square of the Schmidt coefficient.
\end{rem}

In the i.i.d.\ case of $P$,
these quantities are calculated as
\begin{align}
S^*_1(\delta|\overline{P})&= H(P) ,\nonumber \\
S^*_2(\delta|\overline{P})&= 
\min_{0 \,< s \le 1} \frac{s\delta +\psi(s)}{1-s},
\quad
\psi(s)\defeq \log \sum_\omega P(\omega)^s,\Label{19-12}
\end{align}
where 
we use the known value of the left hand side of (\ref{19-11}),
in the calculation (\ref{19-12}).

\begin{rem}\label{8-7-20}
As is discussed in Theorem 3 of Hayashi \cite{ha},
when the limit $\overline{\psi}(s):=\lim_n \frac{1}{n}\log 
\sum_\omega p_n(\omega)^s $ 
and its first and second derivatives $\overline{\psi}'(s)$ and 
$\overline{\psi}''(s)$ exist for 
$s \in (0,1)$,
the relation 
\begin{align}
S^*_2(\delta|\overline{P})=
\min_{0 \,< s \le 1} \frac{s\delta +\overline{\psi}(s)}{1-s}
\end{align}
holds.
\end{rem}

From the above discussion, we find that
changing the order of input distributions of 
KL-divergence causes a completely different asymptotic behavior.

\subsection{Second order asymptotics}
Similar to the variational distance criterion,
in order to more deeply discuss Folklore for source coding,
we need to treat the second order asymptotics. 
For this purpose, we focus on 
the following values:
\begin{align*}
& S^*(\delta,a|\overline{p})\\
\defeq &
\sup_{\{\Psi_n\}}\{
\vliminf \frac{1}{\sqrt{n}}\log \frac{|\Psi_n|}{e^{na}}
|
\vlimsup
\frac{1}{\sqrt{n}}D(p_n\circ \phi_n^{-1}\|p_{U,{\cal M}_n}) \le \delta
\},\\
&S^*_1(\delta,a|\overline{p}) \\
\defeq &
\sup_{\{\Psi_n\}}\{
\vliminf \frac{1}{\sqrt{n}}\log  \frac{|\Psi_n|}{e^{na}}
|
\vlimsup
D(p_{U,{\cal M}_n}\|p_n\circ \phi_n^{-1}) \,< \delta
\}.
\end{align*}
Concerning the first value, the following theorem holds.
\begin{thm}\Label{th91}
Assume that the condition (\ref{8-5-1}) 
and the equation
$\overline{H}(\epsilon,\underline{H}(\overline{p})|\overline{p})=
\underline{H}(\epsilon,\underline{H}(\overline{p})|\overline{p})$
hold.
Define the probability distribution function
$F$ by 
\begin{align*}
\int_0^{\underline{H}(\epsilon,\underline{H}(\overline{p})|\overline{p})}  
F(\,d x) = \epsilon.
\end{align*}
Then, the inequality
\begin{align}
\vliminf\frac{1}{\sqrt{n}}
D(P^n\circ \phi_n^{-1}\|p_{U,{\cal M}_n})
\ge 
\int_{-\infty}^{b}
(b-x)F(\,d x) \Label{8-5-2}
\end{align}
holds, where
$b = \vliminf \frac{1}{\sqrt{n}}\log 
\frac{M_n}{e^{n\underline{H}(\overline{p})}}$.
Furthermore, 
when $\underline{H}
(1-\epsilon,\underline{H}(\overline{p})|\overline{p})= b$,
there exists a sequence of codes $\{\Phi_n\}$ attaining 
the equality (\ref{8-5-2}) and satisfying 
$\lim \varepsilon (\Phi_n)= \epsilon$.
Finally, we remark that the inequality (\ref{8-5-2})
is equivalent to the inequality:
\begin{align}
\vliminf\frac{1}{\sqrt{n}}(H(p_n\circ \phi_n^{-1})
- n \underline{H}(\overline{p}))
\le \int_{0}^{b}
x F(\,d x) + b(1-F(b)). \Label{8-5-5-1}
\end{align}
\end{thm}
Therefore, we obtain
\begin{align*}
S^*(\delta,\underline{H}(P)|\overline{p})
=
\sup_b
\left\{ b \left| \int_{-\infty}^b (b-x)F(\,d x)\le \delta\right.\right\}.
\end{align*}
Concerning the opposite criterion, the following theorem holds.
\begin{thm}\Label{th20}
\begin{align}
S^*_1(\delta,a|\overline{p})
= \underline{H}(1- e^{-\delta},a|\overline{p}).
\end{align}
\end{thm}
In the i.i.d.\ case of $P$,
these quantities are simplified to
\begin{align*}
S^*(\delta,H(P)|\overline{P})
& = \sup_b
\left\{ b \left|
\sqrt{V_P}\int_{-\infty}^{b}\frac{b-x}{\sqrt{2\pi}}e^{-x^2/2}\,d x
\le \delta\right.\right\} \\
S^*_1(\delta,H(P)|\overline{P})
& = 
\sqrt{V_P}\Phi^{-1}(1- e^{-\delta}).
\end{align*}
Especially, when we take the limit $\delta \to 0$,
the relations 
\begin{align*}
S^*(\delta,H(P)|\overline{P})\to - \infty,\quad
S^*_1(\delta,H(P)|\overline{P})\to - \infty
\end{align*}
hold.
On the other hand, 
Theorem \ref{th4} guarantees that
$ R^\ddagger(\epsilon,a|\overline{p})
= \underline{H}(1-\epsilon,a|\overline{p})$, and
$\lim_{\epsilon \to 0} 
\underline{H}(1-\epsilon,a|\overline{P}) = + \infty$.
Thus, 
if a sequence of codes $\Phi_n=({\cal M}_n,\phi_n,\psi_n)$ satisfies 
that
$\varepsilon(\Phi_n)\to 0$,
it does not satisfy 
\begin{align}
\frac{1}{\sqrt{n}}D(p_n\circ \phi_n^{-1}\|p_{U,{\cal M}_n})  \to 0
\Label{19-26}
\end{align}
nor
\begin{align}
D(p_{U,{\cal M}_n}\|p_n\circ \phi_n^{-1})  \to 0.
\Label{19-27}
\end{align}
Therefore, even if we focus on KL-divergence,
if we adopt the criterion (\ref{19-26}) or (\ref{19-27}),
Folklore of source coding does not hold.

Furthermore, combining Theorem \ref{th4}, we obtain
the following corollary.
\begin{cor}
Assume the same assumption as Theorem \ref{th91}.
If the function $\epsilon \mapsto \overline{H}(\epsilon,
\underline{H}(\overline{p})|\overline{p})$
is continuous,
then
\begin{align}
& \inf_{\{\Phi_n\}}
\left. \left\{ \varlimsup \frac{1}{\sqrt{n}}
D(p_n\circ\phi_n^{-1}\|p_{U,{\cal M}_n})
\right |
\varlimsup \varepsilon(\Phi_n) \le \epsilon
\right\} \nonumber \\
\le &
\inf_{\delta}
\left\{  \delta|
S^*(\delta,\overline{H}(\overline{p})|\overline{p})
\ge 
R^{\ddagger}(\epsilon,\overline{H}(\overline{p})
|\overline{p})\right\} \nonumber \\
=& 
\int_{-\infty}^{F^{-1}(1-\epsilon) }
(F^{-1}(1-\epsilon) -x)F(\,d x) ,\Label{9-30-1}\\
& \inf_{\{\Phi_n\}}
\left. \left\{ \varlimsup 
D(p_{U,{\cal M}_n}\|p_n\circ\phi_n^{-1})
\right |
\varlimsup \varepsilon(\Phi_n) \le \epsilon
\right\} \nonumber \\
\le &\inf_{\delta}
\left\{  \delta|
S_1^*(\delta,\overline{H}(\overline{p})|\overline{p})
\ge 
R^{\ddagger}(\epsilon,\overline{H}(\overline{p})
|\overline{p})\right\} \nonumber \\
=& - \log \epsilon.\nonumber 
\end{align}
\end{cor}
In the i.i.d.\ case,
the r. h. s. of (\ref{9-30-1}) equals
\begin{align*}
& \sqrt{V_P}
\int_{-\infty}^{\sqrt{V_P}\Phi^{-1}(1-\epsilon)}
\frac{\sqrt{V_P}\Phi^{-1}(1-\epsilon)-x}{\sqrt{2 \pi}}
e^{-x^2/2}\,d x\\
= &\int_{-\infty}^{\Phi^{-1}(1-\epsilon)}
\frac{\Phi^{-1}(1-\epsilon)-x}{\sqrt{2 \pi}}
e^{-x^2/2}\,d x.
\end{align*}

Finally, we compare the topologies defined by the following limits:
\begin{align}
d(p_n\circ \phi_n^{-1}, p_{U,\cM_n})\to 0 \label{8-6-30}\\
\frac{1}{n}D(p_n\circ \phi_n^{-1}\| p_{U,\cM_n})\to 0 \label{8-6-31}\\
\frac{1}{\sqrt{n}}D(p_n\circ \phi_n^{-1}\| p_{U,\cM_n})\to 0 \label{8-6-32}\\
D(p_n\circ \phi_n^{-1}\| p_{U,\cM_n})\to 0 \label{8-6-33}\\
D( p_{U,\cM_n}\|p_n\circ \phi_n^{-1})\to 0 \label{8-6-34}.
\end{align}
The relations
\begin{align}
&(\ref{8-6-33})
\Rightarrow 
(\ref{8-6-32})
\Rightarrow 
(\ref{8-6-31}), \label{8-6-35} \\
&(\ref{8-6-33})
\Rightarrow 
(\ref{8-6-30})
\Rightarrow 
(\ref{8-6-31}), \label{8-6-36} \\
&(\ref{8-6-34})
\Rightarrow 
(\ref{8-6-30}) \label{8-6-37}
\end{align}
hold.
The relation (\ref{8-6-35}) is trivial, 
the first relation of (\ref{8-6-36}) and the relation (\ref{8-6-37})
is trivial from Pinsker's inequality.
For the second one of (\ref{8-6-36}), see Appendix.

That is, (\ref{8-6-31}) gives the weakest topology among the above 
topologies.
Thus, there is no contradiction, even if 
Folklore for source coding holds in (\ref{8-6-31}), 
but does not hold in (\ref{8-6-32}), (\ref{8-6-34}), or (\ref{8-6-30}).

\section{Markovian case}
Now, we proceed to the Markovian case with irreducible transition matrix
$Q_{j,i}$, where $i$ indicates the input signal and $j$ does the output signal.
When the initial distribution is the stationary distribution $P_i$,
which is the eigen vector of $Q_{j,i}$ with eigen value $1$,
the average $H_n(Q)$ of the normalized likelihood 
can be calculated as
\begin{align*}
& H_n(Q)\\
=& -
\rE_{i_1,\ldots, i_n}
\frac{1}{n}
\log 
Q_{i_n,i_{n-1}} \cdots Q_{i_2,i_1}P_{i_1}\\
=
& -\frac{1}{n}
\sum_{i_{n-1}, i_n}
P_{i_{n-1}}
Q_{i_n,i_{n-1}} 
\log 
Q_{i_n,i_{n-1}} 
+\cdots \\
& +
\sum_{i_1, i_2}
P_{i_{1}}Q_{i_2,i_1} 
\log 
Q_{i_2,i_1} 
+
\sum_{i_1}P_{i_{1}}\log P_{i_{1}}\\
=& 
- \frac{1}{n}\sum_i P_i \log P_i
-\frac{n-1}{n}\sum_{j,i}P_{i}Q_{j,i} \log Q_{j,i}\\
\to & 
H(Q)
:=
-\sum_{j,i}
P_{i}Q_{j,i} 
\log Q_{j,i},
\end{align*}
where 
$\rE_{i_1,\ldots, i_n}$
is the expectation concerning the distribution
$Q_{i_n,i_{n-1}} \cdots Q_{i_2,i_1}P_{i_1}$.

In oder to treat 
the limit distribution of 
the normalized likelihood, 
we calculate the second cumulant as
\begin{align*}
& 
\rE_{i_1,\ldots, i_n}
\left(
\frac{
-\log 
Q_{i_n,i_{n-1}} 
\cdots Q_{i_2,i_1}P_{i_1}
-n H_n(Q)
}{\sqrt{n}}
\right)^2\\
=&
\rE_{i_1,\ldots, i_n}
\left(
\frac{
X(i_{n},i_{n-1})
+ 
\cdots 
+
X(i_{2},i_1)
+
Y(i_1)
}{\sqrt{n}}
\right)^2\\
=&
\rE_{i_1,\ldots, i_n}
\frac{1}{n}
\Bigl(
X(i_{n},i_{n-1})^2
+\cdots +X(i_{2},i_1)^2 \\
&+
Y(i_1)^2
+ 2X(i_{n},i_{n-1})X(i_{n-1},i_{n-2})
+\cdots \\
& +
2X(i_{3},i_{2})X(i_{2},i_{1})
+2 X(i_{2},i_{1})Y(i_1)
\Bigr)\\
\to & V(Q),
\end{align*}
where
$X(i_{k+1},i_k)
:=-\log Q_{i_{k+1},i_k}-H(Q)$, 
$Y(i_1):=P_{i_1}-H(P)$, and 
\begin{align*}
&V(Q)\\
:=&
\sum_{j,i}Q_{j,i}P_i
(-\log 
Q_{j,i}-H(Q))^2 \\
& + 
2
\sum_{k,j,i}Q_{k,j}Q_{j,i}P_i
(-\log Q_{k,j}-H(Q))
(-\log Q_{j,i}-H(Q)).
\end{align*}
The limit of the third cumulant is calculated as
\begin{align*}
& 
\rE_{i_1,\ldots, i_n}
\left(
\frac{
-\log 
Q_{i_n,i_{n-1}} 
\cdots Q_{i_2,i_1}P_{i_1}
-n H_n(Q)
}{\sqrt{n}}
\right)^3\\
=&
\rE_{i_1,\ldots, i_n}
\left(
\frac{
X(i_{n},i_{n-1})
+ 
\cdots 
+
X(i_{2},i_1)
+
Y(i_1)
}{\sqrt{n}}
\right)^3\\
=&
\rE_{i_1,\ldots, i_n}
\frac{1}{n\sqrt{n}}
\Bigl(
X(i_{n},i_{n-1})^3
+\cdots +X(i_{2},i_1)^3 
+Y(i_1)^3 \\
& + 3 (X(i_{n},i_{n-1})^2 X(i_{n-1},i_{n-2})
+\cdots \\&\quad+
X(i_{3},i_{2})^2 X(i_{2},i_{1})
+ X(i_{2},i_{1})^2 Y(i_1))\\
& + 3 (X(i_{n},i_{n-1}) X(i_{n-1},i_{n-2})^2
+\cdots \\& \quad +
X(i_{3},i_{2}) X(i_{2},i_{1})^2
+ X(i_{2},i_{1}) Y(i_1))^2\\
& + 2 (X(i_{n},i_{n-1}) X(i_{n-1},i_{n-2})X(i_{n-2},i_{n-3})
+\cdots \\& \quad+
X(i_{4},i_{3}) X(i_{3},i_{2}) X(i_{2},i_{1})
+ X(i_{3},i_{2}) X(i_{2},i_{1}) Y(i_1))
\Bigr)\\
\to & 0.
\end{align*}
Similarly, for $n \ge 3$, the $n$-th
cumulant goes to $0$ because the 
numerator is linear for $n$ while the
denominator is a higher term for $n$.
Thus, the limit distribution of the normalized likelihood
is equal to the normal distribution 
with average $H(Q)$ and
the variance $V(Q)$.
Hence, concerning the first order asymptotics, 
we have
\begin{align}
\overline{H}(0 |\overline{Q})
=\underline{H}_+(1 |\overline{Q})=H(Q),
\end{align}
where
$\overline{Q}\defeq \{Q^n\}$ and 
$Q^n_{i_n,\ldots,i_1}\defeq 
Q_{i_n,i_{n-1}} \cdots Q_{i_2,i_1}P_{i_1}$.
concerning the second order asymptotics, 
we have
\begin{align}
\underline{H}(\epsilon,H(Q) |\overline{Q})
=\overline{H}(\epsilon,H(Q) |\overline{Q})
= \sqrt{V(Q)}\Phi^{-1}(\epsilon) \Label{25-a}.
\end{align}

Next, we consider the case 
the initial distribution is the different from 
the stationary distribution $P_i$.
In this case, the distribution of the $n$-th data
exponentially approaches to the stationary distribution $P_i$ \cite{DZ}.
Hence, 
the limit distribution of the normalized likelihood
is equal to the normal distribution 
with average $H(Q)$ and
the variance $V(Q)$.
Therefore, 
Folklore for source coding does not hold 
for the topology (\ref{8-6-32}), (\ref{8-6-34}), or (\ref{8-6-30})
in the Markovian case as in the i.i.d. case.

Further, 
by using Remark \ref{8-7-20},
$S^*_2(\delta|\overline{P})$ is calculated as follows.
In the Markovian case, 
$\overline{\psi}(s)=
\log \sum_\omega Q_{j,i}^sP_{s;i}$,
where the distribution $P_{s;i}$ consists of eigen vectors of
the matrix $Q_{j,i}^s$ (Section 3 of Dembo \& Zeitouni \cite{DZ}).
Hence, we obtain 
\begin{align}
S^*_2(\delta|\overline{P})= 
\min_{0 \,< s \le 1} \frac{s\delta +\overline{\psi}(s)}{1-s}.
\end{align}

\section{Universal fixed-length source coding and 
universal intrinsic randomness}
In this section, we focus only on the independent and 
identical information source.
In this case, as was shown by Csisz\'{a}r and K\"{o}rner\cite{CK},
there exists a fixed-length source code 
that attains the first order optimal rate and does not depend on 
the probability distribution of the information source
while we proved the existence of a code attaining the optimal bound
depending on the distribution.
Such a code is called universal, and is an important topic
in information theory.
Indeed, information spectrum method can apply any sequence of 
information source, but gives a code depending on this information source.
In contrast, universal code assumes 
on the independent and identical information source, (or Markovian source),
but depends only on the coding rate not on the information source.
As is stated in the following theorem,
there exists a universal fixed-length source code 
attaining the second order optimal rate.
\begin{thm}\Label{21-8}
Assume that $|\Omega|$ is a finite number $d$,
then there exists a fixed-length source code $\Phi_n$ 
on $\Omega^n$ such that 
\begin{align}
\lim \frac{1}{\sqrt{n}}\frac{|\Phi_n|}{e^{na}}= b\Label{20-2}
\end{align}
and 
\begin{align}
\lim \varepsilon_{P^n}(\Phi_n) =
\left\{ 
\begin{array}{cl}
0 & H(P) \,< a \\
1- \Phi(\frac{b}{\sqrt{V_P}}) & H(P) = a.
\end{array}
\right.
\Label{20-3}
\end{align}
\end{thm}
The error probability of the universal fixed-length source
code had not been discussed 
when the rate equaled the entropy of the information source.
But, this theorem clarifies asymptotic behavior of the error 
probability in such a special case
by treating the second order asymptotics.

Concerning intrinsic randomness,
while Oohama and Sugano \cite{OS} proved
that there exists an operation universally attaining 
the first order optimal rate,
we can also prove the existence of 
a universal operation achieving 
the second order optimal rate.
\begin{thm}\Label{21-9}
Assume that $|\Omega|$ is a finite number $d$,
then there exists an operation $\Psi_n$ 
on $\Omega^n$ such that 
\begin{align}
\lim \frac{1}{\sqrt{n}}\frac{|\Phi_n|}{e^{na}}= b
\end{align}
and 
\begin{align}
\lim \varepsilon_{P^n}(\Psi_n) =
\left\{ 
\begin{array}{cl}
0 & H(P) \,> a \\
\Phi(\frac{b}{\sqrt{V_P}}) & H(P) = a.
\end{array}
\right.\Label{20-5}
\end{align}
\end{thm}
\section{Proof of theorems}
First, we give proofs of 
Theorems \ref{th1} and \ref{th3}, which are partially known.
Following these proofs, we give our proof of 
Theorem \ref{th4}, which is the main result of this paper.
This is because the former are  preliminaries to our proof of 
Theorem \ref{th4}.
After these proofs, we give proofs of 
Theorems \ref{thm10}--\ref{th20}.

\subsection{Proof of Theorem \ref{th1}}
\begin{lem}Han \cite[Lemma 1.3.1]{Han1}\Label{le3}
For any integer $M_n$,
there exists a code $\Phi_n$ satisfying
\begin{align}
1- \varepsilon(\Phi_n)
\ge p_n \{ p_n (\omega) \,> \frac{1}{M_n}\}, \quad
|\Phi_n|  \le M_n. 
\end{align}
\end{lem}
\begin{lem}
Han \cite[Lemma 1.3.2]{Han1}\Label{le4}
Any integer $M_n'$ and any code $\Phi_n$ satisfy the
following condition:
\begin{align*}
1- \varepsilon(\Phi_n)
\le p_n \{ p_n (\omega) \,> \frac{1}{M_n'}\}
+ \frac{|\Phi_n|}{M_n'}.
\end{align*}
\end{lem}
By using these lemmas and 
the following expressions of the quantities
$R(1-\epsilon|\overline{p}),R^\dagger(1-\epsilon|\overline{p})$ and
$R^\ddagger(1-\epsilon|\overline{p})$,
we will prove Theorem \ref{th1}.
\begin{align*}
R(1-\epsilon|\overline{p})
&= 
\inf_{\{\Phi_n\}}
\{
\vlimsup \frac{1}{n} \log |\Phi_n| |
\vliminf 1- \varepsilon(\Phi_n) \ge \epsilon\},\\
R^\dagger(1-\epsilon|\overline{p})
&= 
\inf_{\{\Phi_n\}}
\{
\vlimsup \frac{1}{n} \log |\Phi_n| |
\vlimsup 1- \varepsilon(\Phi_n)\ge \epsilon\},\\
R^\ddagger(1-\epsilon|\overline{p})
&= 
\inf_{\{\Phi_n\}}
\{
\vliminf \frac{1}{n} \log |\Phi_n| |
\vliminf 1- \varepsilon(\Phi_n) \ge \epsilon\}.
\end{align*}
\noindent{\em Proof of direct part:}\quad
For any real number $a > \overline{H}(\epsilon|\overline{p})$,
by applying Lemma \ref{le3} to the case of $M_n = e^{na}$,
we can show
\begin{align}
\vliminf p_n \{ p_n (\omega) > \frac{1}{M_n}\}
= 
\vliminf p_n \{ - \frac{1}{n} \log p_n (\omega) < a \}
\ge \epsilon, \Label{eq14}
\end{align}
which implies that
$a \ge R(1-\epsilon|\overline{p})$.
Thus, we obtain
\begin{align*}
\overline{H}(\epsilon|\overline{p})\ge R(1-\epsilon|\overline{p}).
\end{align*}
By replacing the limit $\vliminf$ in (\ref{eq14}) by $\vlimsup$,
we can show
\begin{align*}
\underline{H}(\epsilon|\overline{p})\ge R^\dagger(1-\epsilon|\overline{p}).
\end{align*}
Finally, by choosing 
$M_n$ satisfying
\begin{align*}
&\vliminf p_n \{ -\frac{1}{n}\log p_n (\omega) < \frac{1}{n}\log M_n\}
\ge \epsilon\\
&\vliminf \frac{1}{n} \log M_n = a > \underline{H}(\epsilon|\overline{p}),
\end{align*}
we can prove
\begin{align*}
\underline{H}(\epsilon|\overline{p})\ge R^\ddagger(1-\epsilon|\overline{p}).
\end{align*}
The direct part of (\ref{8-5-22}) and (\ref{8-5-23})
can be proved by replacing $\ge \epsilon $ by $> \epsilon $
in the above proof.
\endproof

\noindent{\em Proof of converse part:}\quad
First, we prove
\begin{align}
\overline{H}(\epsilon|\overline{p})\le R(1-\epsilon|\overline{p})
\Label{eq18}.
\end{align}
Assume that $a\defeq \vlimsup \frac{1}{n}\log |\Phi_n|$,
$\vliminf 1- \varepsilon (\Phi_n) \ge \epsilon$.
For any real number $\delta\,> 0$,
we apply Lemma \ref{le4} to the case of $M_n'= e^{n(a+ \delta)}$.
Then, we obtain
\begin{align}
p_n \{ -\frac{1}{n}\log p_n(\omega)
< a+ \delta\}
\ge 1- \varepsilon(\Phi_n) - \frac{|\Phi_n|}{e^{n(a+ \delta)}}.
\Label{eq19}
\end{align}
Taking the limit $\vliminf$, we can show
\begin{align*}
\vliminf p_n \{ -\frac{1}{n}\log p_n(\omega)
< a+ \delta\} \ge \epsilon.
\end{align*}
From this relation, we obtain 
$a + \delta \ge \overline{H}(\epsilon|\overline{p})$,
which implies (\ref{eq18}).

Similarly, taking the limit $\vlimsup$ at (\ref{eq19}),
we can prove
\begin{align*}
\underline{H}(\epsilon|\overline{p})\le R^\dagger(1-\epsilon|\overline{p}).
\end{align*}
Finally, we focus on a subsequence $n_k$
satisfying $a \defeq \vliminf \frac{1}{n}\log |\Phi_n|=
\lim_k \frac{1}{n_k}\log |\Phi_{n_k}|$.
By using (\ref{eq19}),
we obtain
\begin{align*}
&\vliminf p_n \{ -\frac{1}{n}\log p_n(\omega)
< a- \delta\} \\
\le &\lim_k p_{n_k} \{ -\frac{1}{n_k}\log p_{n_k}(\omega)
< a- \delta\}
\le \lim_k 1- \varepsilon(\Phi_{n_k}).
\end{align*}
Taking account into the above discussions,
we can prove
\begin{align*}
\underline{H}(\epsilon|\overline{p})
\le R^\ddagger(1-\epsilon|\overline{p}).
\end{align*}
Similarly, the converse part of (\ref{8-5-22}) and (\ref{8-5-23})
can be proved by replacing $\ge \epsilon $ by $> \epsilon $ in the above proof.
\endproof

\subsection{Proof of Theorem \ref{th3}}
\begin{lem}
Han \cite[Lemma 2.1.1]{Han1}\Label{le1}
For any integers $M_n'$ and $M_n$,
there exists an operation $\Psi_n= ({\cal M}_n,\phi_n)$ satisfying 
\begin{align}
\varepsilon (\Psi_n)  \le 
p_n \{ p_n (\omega) > \frac{1}{M_n'}\} +\frac{M_n}{M'_n},~
|\Psi_n| = M_n .
\end{align}
\end{lem}
\begin{lem}
Han \cite[Lemma 2.1.2]{Han1}\Label{le2}
Any integer $M_n'$ and any operation $\Psi_n$ satisfy
\begin{align*}
\varepsilon (\Psi_n)  \ge
p_n \{ p_n (\omega) > \frac{1}{M_n'}\} -\frac{M_n'}{|\Psi_n|}.
\end{align*}
\end{lem}

By using these lemmas, we prove Theorem \ref{th3}.

\noindent{\em Proof of direct part:}\quad
For any real numbers $a < \underline{H}(\epsilon|\overline{p})$
and $\delta \,>0$,
we apply Lemma \ref{le1} to the case of 
$M_n = e^{n(a - \delta)},M_n' = e^{na}$ as follows:
\begin{align}
& \vlimsup p_n \{ p_n (\omega) > \frac{1}{M_n'}\} \nonumber \\
= &
\vlimsup p_n \{ - \frac{1}{n} \log p_n (\omega) < a \}
\,< \epsilon \Label{eq4}.
\end{align}
Since $\frac{M_n}{M'_n} \to 0 $,
we obtain $\vlimsup \varepsilon (\Psi_n) \,< \epsilon$,
which implies that
$a -\delta \le S(\epsilon|\overline{p})$.
Thus, the inequality
\begin{align*}
\underline{H}(\epsilon|\overline{p})\le S(\epsilon|\overline{p})
\end{align*}
holds.
Moreover, by replacing the limit in 
(\ref{eq4}) by $\vliminf$,
we can prove 
\begin{align*}
\overline{H}(\epsilon|\overline{p})\le S^\dagger(\epsilon|\overline{p})
\end{align*}
Finally, 
by choosing $M_n'$ satisfying
\begin{align*}
&\vlimsup p_n \{ -\frac{1}{n}\log p_n (\omega) < \frac{1}{n}\log M_n'\}
< \epsilon\\
&\vlimsup \frac{1}{n} \log M_n' = a < \overline{H}(\epsilon|\overline{p}),
\end{align*}
we can prove 
\begin{align*}
\overline{H}(\epsilon|\overline{p})\le S^\ddagger(\epsilon|\overline{p}).
\end{align*}
The direct part of (\ref{8-5-16}) can be proved by replacing 
$< \epsilon$ by $\le \epsilon$ in the above proof.
\endproof

\noindent{\em Proof of converse part:}\quad
First, we prove
\begin{align}
\underline{H}(\epsilon|\overline{p})\ge S(\epsilon|\overline{p})\Label{eq8}.
\end{align}
Assume that
$a\defeq \vliminf \frac{1}{n}\log |\Psi_n|$,
$\vlimsup \epsilon (\Psi_n) \,< \epsilon$.
For any real number $\delta\,> 0$,
we apply Lemma \ref{le2} to the case of
$M_n'= e^{n(a- \delta)}$.
Then, we obtain
\begin{align}
p_n \{ -\frac{1}{n}\log p_n(\omega)
< a- \delta\}
\le \varepsilon(\Psi_n) + \frac{e^{n(a- \delta)}}{|\Psi_n|}.
\Label{eq9}
\end{align}
Taking the limit $\vlimsup$, we can show that
\begin{align*}
\vlimsup p_n \{ -\frac{1}{n}\log p_n(\omega)
< a- \delta\} < \epsilon.
\end{align*}
Thus, we obtain 
$a - \delta \le \underline{H}(\epsilon|\overline{p})$,
which implies (\ref{eq8}).

Similarly, by taking the limit $\vliminf$ at the inequality (\ref{eq9}),
we obtain 
\begin{align*}
\overline{H}(\epsilon|\overline{p})\ge S^\dagger(\epsilon|\overline{p}).
\end{align*}
Moreover, by focusing on
a subsequence $n_k$ satisfying
$a \defeq \vlimsup \frac{1}{n}\log |\Psi_n|=
\lim_k \frac{1}{n_k}\log |\Psi_{n_k}|$,
we can show the following relations from (\ref{eq9}):
\begin{align*}
&\vliminf p_n \{ -\frac{1}{n}\log p_n(\omega)
< a- \delta\} \\
\le &\lim_k p_{n_k} \{ -\frac{1}{n_k}\log p_{n_k}(\omega)
< a- \delta\}\le \lim_k \varepsilon(\Psi_{n_k}),
\end{align*}
which implies that
\begin{align*}
\overline{H}(\epsilon|\overline{p})\ge S^\ddagger(\epsilon|\overline{p}).
\end{align*}
Similarly, the converse part of (\ref{8-5-16}) can be proved by replacing 
$< \epsilon$ by $\le \epsilon$ in the above proof.
\endproof

\subsection{Proof of Theorem \ref{th4}}
For any real number $b > \overline{H}(\epsilon,a|\overline{p})$,
by applying Lemma \ref{le3} to the case of $M_n = e^{na+\sqrt{n}b}$,
we can show
\begin{align*}
& \vliminf p_n \{ p_n (\omega) > \frac{1}{M_n}\} \nonumber\\
= &
\vliminf p_n \{ - \frac{1}{n} \log p_n (\omega) < a+ \frac{b}{\sqrt{n}} \}
\ge \epsilon,
\end{align*}
which implies 
$b \ge R(1-\epsilon,a|\overline{p})$.
Thus, we obtain
\begin{align*}
\overline{H}(\epsilon,a|\overline{p})\ge R(1-\epsilon,a|\overline{p}).
\end{align*}
Similarly to Proof of Theorem \ref{th1},
we can show
\begin{align*}
\underline{H}(\epsilon,a|\overline{p})\ge R^\dagger(1-\epsilon,a|\overline{p})
,\quad
\underline{H}(\epsilon,a|\overline{p})\ge 
R^\ddagger(1-\epsilon,a|\overline{p}).
\end{align*}

Next, we prove 
\begin{align}
\overline{H}(\epsilon,a|\overline{p})\ge R(1-\epsilon,a|\overline{p})
\Label{eq118}.
\end{align}
Assume that
$b\defeq \vlimsup \frac{1}{n}\log \frac{|\Phi_n|}{e^{na}}$,
$\vliminf 1- \epsilon (\Phi_n) \ge \epsilon$.
For any real number $\delta\,> 0$,
we apply Lemma \ref{le4} to the case of
$M_n'= e^{na + \sqrt{n}(b+ \delta)}$.
Then, we obtain
\begin{align*}
p_n \{ -\frac{1}{n}\log p_n(\omega)
< a+ \frac{b+\delta}{\sqrt{n}} \}
\ge 1- \varepsilon(\Phi_n) - \frac{|\Phi_n|}{e^{na + \sqrt{n}(b+ \delta)}}.
\end{align*}
Taking the limit $\vliminf$, we obtain
\begin{align*}
\vliminf p_n \{ -\frac{1}{n}\log p_n(\omega)
< a+  \frac{b+\delta}{\sqrt{n}}\} \ge \epsilon,
\end{align*}
which implies 
$b + \delta \ge \overline{H}(\epsilon,a|\overline{p})$.
Thus, we obtain (\ref{eq118}).
Therefore, similar to our proof of Theorem \ref{th1},
we can show
\begin{align*}
\underline{H}(\epsilon,a|\overline{p})\ge 
R^\dagger(1-\epsilon,a|\overline{p}), 
\quad
\underline{H}(\epsilon,a|\overline{p})\ge 
R^\ddagger(1-\epsilon,a|\overline{p}).
\end{align*}

Next, we prove
\begin{align}
\underline{H}(\epsilon,a|\overline{p})\le S(\epsilon|\overline{p})
\Label{41}.
\end{align}
For any real numbers $b < \underline{H}(\epsilon,a|\overline{p})$
and $\delta \,>0$,
we apply Lemma \ref{le1} to the case of
$M_n = e^{na + \sqrt{n}(b - \delta)},M_n' = e^{na + \sqrt{n}b}$.
Since
\begin{align*}
& \vlimsup p_n \{ p_n (\omega) > \frac{1}{M_n'}\} \nonumber \\
= &
\vlimsup p_n \{ - \frac{1}{n} \log p_n (\omega) < a + \frac{b}{\sqrt{n}}
\}
\,< \epsilon 
\end{align*}
and $\frac{M_n}{M'_n} \to 0 $,
we obtain $\vlimsup \epsilon (\Psi_n) \,< \epsilon$ which implies 
$a -\delta \le S(\epsilon|\overline{p})$.
Thus, we obtain (\ref{41}).

Similar to our proof of Theorem \ref{th3},
we can show
\begin{align*}
\overline{H}(\epsilon,a|\overline{p})\le 
S^\dagger(\epsilon,a|\overline{p}),
\quad
\overline{H}(\epsilon,a|\overline{p})\le S^\ddagger(\epsilon,a|\overline{p}).
\end{align*}

Finally, we prove 
\begin{align}
\underline{H}(\epsilon,a|\overline{p})
\ge S(\epsilon|\overline{p})\Label{eq88}.
\end{align}
Assume that
$b\defeq \vliminf \frac{1}{n}\log \frac{|\Psi_n|}{e^{na}}$ and 
$\vlimsup \epsilon (\Psi_n) \,< \epsilon$.
For any real number $\delta\,> 0$,
we apply Lemma \ref{le2} to the case of
$M_n'= e^{na + \sqrt{n}(b- \delta)}$.
Then, the inequality
\begin{align*}
p_n \{ -\frac{1}{n}\log p_n(\omega)
< a+ \frac{b- \delta}{\sqrt{n}}\}
\le \varepsilon(\Psi_n) + \frac{e^{na + \sqrt{n}(b- \delta)}}{|\Psi_n|}
\end{align*}
holds.
Taking the limit $\vlimsup$, we obtain
\begin{align*}
\vlimsup p_n \{ -\frac{1}{n}\log p_n(\omega)
< a+ \frac{b - \delta}{\sqrt{n}}\} < \epsilon,
\end{align*}
which implies 
$b - \delta \le \underline{H}(\epsilon,a|\overline{p})$.
Thus, the relation 
(\ref{eq88}) holds.

Similar to our proof of Theorem \ref{th3},
the inequalities
\begin{align*}
\overline{H}(\epsilon,a|\overline{p})\ge 
S^\dagger(\epsilon,a|\overline{p}),
\quad
\overline{H}(\epsilon,a|\overline{p})\ge 
S^\ddagger(\epsilon,a|\overline{p})
\end{align*}
are proved.
\endproof
\subsection{Proof of Theorem \ref{thm10}}
We define the subset ${\cal M}_n'$ of ${\cal M}_n$ as
\begin{align*}
{\cal M}_n'\defeq \{ i \in {\cal M}_n|
\psi_n(i) \in \phi_n^{-1}( i)\}.
\end{align*}
Since the relation $\phi_n^{-1}( i) \cap \phi_n^{-1}(j)= \emptyset $
holds for any distinct integers $i,j$,
the map $\psi_n$ is injective on ${\cal M}_n'$.
Thus,
$p_n$ can be regarded as a probability distribution on
 ${\cal M}_n'\cup (\Omega_n \setminus \psi_n({\cal M}_n'))
\subset {\cal M}_n\cup (\Omega_n \setminus \psi_n({\cal M}_n'))
$.
Similarly, 
$p_n\circ \phi_n^{-1}$ also can be regarded as
a probability distribution on
${\cal M}_n\subset {\cal M}_n\cup (\Omega_n \setminus \psi_n({\cal M}_n'))
$.

Then, the relation 
\begin{align*}
d(p_n, p_{U,{\cal M}_n'})\le
d(p_n, p_{U,{\cal M}_n})
\end{align*}
holds.
The definition of $\delta(p_n)$ guarantees that
\begin{align*}
\delta(p_n) \le d(p_n, p_{U,{\cal M}_n'}).
\end{align*}
The axiom of distance yields that
\begin{align*}
d(p_n, p_{U,{\cal M}_n}) \le d(p_n,p_n\circ \phi_n^{-1})
+ d(p_n\circ \phi_n^{-1}, p_{U,{\cal M}_n}).
\end{align*}
Furthermore, the quantity 
$\varepsilon (\Phi_n)$ has another expression:
$\varepsilon (\Phi_n)=p_n(\Omega_n \setminus \psi_n({\cal M}_n'))$.

Since 
the set $(\Omega_n \setminus \psi_n({\cal M}_n'))$
coincides with the set of the element of 
${\cal M}_n\cup (\Omega_n \setminus \psi_n({\cal M}_n'))$ 
such that the probability 
$p_n$ is greater than the probability $p_n\circ \phi_n^{-1}$,
the equation
\begin{align*}
d(p_n,p_n\circ \phi_n^{-1})
=p_n(\Omega_n \setminus \psi_n({\cal M}_n'))= 
\varepsilon (\Phi_n)
\end{align*}
holds.

Combining the above relations,
we obtain 
\begin{align*}
\delta(p_n) \le \varepsilon (\Phi_n) + \varepsilon (\Psi_n).
\end{align*}
\endproof

\subsection{Proof of Theorem \ref{th5}}
First, we construct a sequence of codes $\Phi_n= ({\cal M}_n,\phi_n,\psi_n)$ 
satisfying (\ref{35}) and (\ref{36})
as follows.
We assume that
$S_n(a,b)\defeq
\{ - \frac{1}{n} \log p_n (\omega) < a+ \frac{b}{\sqrt{n}} \}$,
$\tilde{M}_n\defeq |S_n(a,b) |$ 
and denote
the one-to-one map from
$S_n(a,b)$ to ${\cal \tilde{M}}_n\defeq
\{1, \ldots, \tilde{M}_n\}$ by $\tilde{\phi}_n$.

Then, the inequality $\tilde{M}_n \le e^{na +\sqrt{n}b}$ holds.
Next, we 
define $\epsilon_n \defeq p_n (S_n(a,b))$ and
focus on 
the probability distribution
$\hat{p}_n(\omega)\defeq \frac{p_n(\omega)}{1-\epsilon_n}$
on $S_n(a,b)^c$.
Then, we apply Lemma \ref{le1} to the case of
$M_n'=\hat{M}_n'\defeq (1-\epsilon_n) e^{na +\sqrt{n}(b+2\gamma_n)}$,
$M_n=\hat{M}_n\defeq (1-\epsilon_n) e^{na +\sqrt{n}(b+\gamma_n)}$,
$\gamma_n= 1/n^{1/4}$, and
denote the transformation satisfying the condition of
Lemma \ref{le1} 
by $\hat{\phi}_n$,
where
the range of $\hat{\phi}_n$ is 
$\{\tilde{M}_n +1, \ldots, \hat{M}_n+ \tilde{M}_n\}$.
Half of the variational distance between
$ \hat{p}_n \circ \hat{\phi}_n^{-1}$
and the uniform distribution is less than
\begin{align}
\hat{p}_n 
\{ - \frac{1}{n} \log p_n (\omega) < a+ \frac{b+2\gamma_n}{\sqrt{n}} \}
+ e^{-\sqrt{n}\gamma_n}.
\end{align}

Next, we define a code $\Phi_n=({\cal M}_n,\phi_n,\psi_n)$ with
the size $M_n=  e^{na +\sqrt{n}(b+\gamma_n)}$ as follows.
The encoding $\phi_n$ is defined by 
$\tilde{\phi}_n$ and $\hat{\phi}_n$.
The decoding $\psi_n$ is defined as 
the inverse map on the subset ${\cal \tilde{M}}_n$ of
${\cal M}_n$,
and is defined as an arbitrary map
on the compliment set ${\cal \tilde{M}}_n^c$.
Since
\begin{align}
1- \varepsilon(\Phi_n) \ge
p_n (S_n (a,b)) = \epsilon_n ,
\end{align}
we obtain the first inequality of (\ref{35}).

Since the variational distance equals the sum of 
that on the range of $\hat{\phi}_n$ 
and that on the compliment set of the range,
$\varepsilon(\Psi_n)$ can be evaluated as follows:
\begin{align*}
&\varepsilon(\Psi_n) \\
\le 
&(1-\epsilon_n)
\left(\hat{p}_n 
\{ - \frac{1}{n} \log p_n (\omega) < a+ \frac{b+2\gamma_n}{\sqrt{n}} \}
+ e^{-\sqrt{n}\gamma_n}\right)\\
&+ p_n (S_n (a,b)) \\
=& 
(1-\epsilon_n)e^{-\sqrt{n}\gamma_n}
+p_n (S_n (a,b))\\
&+ 
p_n 
(S_n (a,b)^c
\cap
\{ - \frac{1}{n} \log p_n (\omega) < a+ \frac{b+2\gamma_n}{\sqrt{n}} \})\\
= &
(1-\epsilon_n)e^{-\sqrt{n}\gamma_n}
+p_n \{ - \frac{1}{n} \log p_n (\omega) < a+ \frac{b+2\gamma_n}{\sqrt{n}} \}
\\
\to & \epsilon,
\end{align*}
where we use the relation 
$S_n (a,b)\subset 
\{ - \frac{1}{n} \log p_n (\omega) < a+ \frac{b+2\gamma_n}{\sqrt{n}} \}$.
Since the definition of $M_n$ guarantees the condition (\ref{36}),
the proof is completed.
\endproof

\subsection{Proof of Theorem \ref{th9}}
\noindent{\em Proof of inequality (\ref{8-5-2-1}):}
\begin{lem}\Label{8-6-1}
The following relation holds for any operation $({\cal M}_n,\phi_n)$:
\begin{align*}
& H(p_n\circ \phi_n^{-1})\\
\le &
H(M_n,p_n) \\
&+ p_n\{ p_n (\omega) \le \frac{1}{M_n} \}
(\log M_n - \log p_n\{ p_n (\omega) \le \frac{1}{M_n} \}),
\end{align*}
where
\begin{align}
H(M_n,p_n) \defeq - \sum_{ p_n (\omega) \,> \frac{1}{M_n}}
p_n (\omega) \log p_n (\omega).
\end{align}
\end{lem}
\begin{proof}
Define the set ${\cal M}_n'$ 
and the map $\phi_n'$ from
$\Omega_n$ to ${\cal M}_n'$ as follows:
\begin{align}
{\cal M}_n' &\defeq {\cal M}_n \cup \{ p_n (\omega) \,> \frac{1}{M_n}\}, 
\\
\phi_n'(\omega) &\defeq 
\left\{
\begin{array}{ll}
\phi_n(\omega) & p_n (\omega) \le \frac{1}{M_n} \\
\omega & p_n (\omega) \,> \frac{1}{M_n}.
\end{array}
\right. 
\end{align}
Since 
\begin{align*}
& -\sum_{i=1}^{M_n} p_n\circ {\phi_n'}^{-1}(i)
\log p_n\circ {\phi_n'}^{-1}(i) \\
\le & p_n\{ p_n (\omega) \le \frac{1}{M_n} \}
(\log M_n - \log p_n\{ p_n (\omega) \le \frac{1}{M_n} \}),
\end{align*}
the inequality
\begin{align*}
& H(p_n\circ {\phi_n'}^{-1})\\
\le &
H(M_n,p_n) \\
& + p_n\{ p_n (\omega) \le \frac{1}{M_n} \}
(\log M_n - \log p_n\{ p_n (\omega) \le \frac{1}{M_n} \})
\end{align*}
holds.
When the map $\phi_n''$ from ${\cal M}_n'$ to ${\cal M}_n$
is defined by
\begin{align*}
\phi_n''(\omega)\defeq 
\left\{
\begin{array}{ll}
\omega & \omega \in {\cal M}_n \\
\phi_n(\omega) & \omega \in  \{p_n (\omega) \,> \frac{1}{M_n}\},
\end{array}
\right.
\end{align*}
the relation $\phi_n= \phi_n''\circ \phi_n'$ holds.
Thus, $\phi_n^{-1}= {\phi_n'}^{-1}\circ {\phi_n''}^{-1}$.
Generally, any map $f$ and any distribution $Q$ satisfies
\begin{align*}
&H(Q\circ f^{-1})
=- \sum_y \sum_{x:y=f(x)}Q(x) \log (\sum_{x':y=f(x')}Q(x'))\\
\le &
- \sum_y \sum_{x:y=f(x)}Q(x) \log Q(x)
= H(Q) .
\end{align*}
Hence, 
\begin{align*}
H(p_n\circ \phi_n^{-1})
= H(p_n\circ {\phi_n'}^{-1}\circ {\phi_n''}^{-1})
\le H(p_n\circ {\phi_n'}^{-1}).
\end{align*}
Therefore, the proof is completed.
\end{proof}
We define the probability distribution function
$F_n$ on the real numbers
$\real$ as:
\begin{align}
F_n(x)\defeq p_n\{ -\frac{1}{n} \log p_n (\omega) \,< x \}
\end{align}
for a probability distribution $p_n$.
Then,
the relation 
\begin{align}
\frac{1}{n}H(M_n,p_n)=
\int_{0}^{\frac{1}{n}\log M_n}x F_n(\,d x)\Label{8-13-1}
\end{align}
holds.
Thus, Lemma \ref{8-6-1} yields the inequality
\begin{align*}
& \frac{1}{n}H(p_n\circ \phi_n^{-1})\\
\le & \int_{0}^{\frac{1}{n}\log M_n}x F_n(\,d x)\\
&+
\frac{1}{n}
p_n\{ p_n (\omega) \le \frac{1}{M_n} \}
(\log M_n - \log p_n\{ p_n (\omega) \le \frac{1}{M_n} \}).
\end{align*}
Taking the limit, we obtain (\ref{8-5-5}).
\endproof

\noindent{\em Proof of the existence part:}\quad
\begin{lem}\Label{le10}
Han\cite[Equation (2.2.4)]{Han1}
For any integers $M_n$ and ${M'}_n$,
there exists an operation $\Psi_n$ such that
\begin{align*}
&D(p_n \circ \psi^{-1}\|p_{U,{\cal M}_n})\nonumber  \\ 
\le & \log M_n \left(
\frac{{M'}_n}{M_n}+ \frac{1}{{M'}_n}
+ p_n \left\{
p_n (\omega) \,> \frac{1}{M_n}\right\} 
\right),\\
|\Psi_n|= & M_n.
\end{align*}
\end{lem}
\begin{rem}
Han \cite{Han1} derived the above inequality
in his proof of Proposition \ref{v-v}.
\end{rem}
In the following, by using Lemma \ref{le10}, 
we construct the code $\Phi_n= ({\cal M}_n,\phi_n,\psi_n)$
satisfying the equality of (\ref{8-5-5}) and
$\vliminf \varepsilon (\Phi_n)= \epsilon$ as follows.
Assume that $S_n(a)\defeq
\{ - \frac{1}{n} \log p_n (\omega) < a \}$,
$\tilde{M}_n\defeq |S_n(a) |$ 
and let $\tilde{\phi}_n$ be the one-to-one map
from $S_n(a)$ to ${\cal \tilde{M}}_n\defeq
\{1, \ldots, \tilde{M}_n\}$.
Then, we can prove that $\tilde{M}_n \le e^{na}$.
Moreover, 
we let $\hat{\phi}_n$ be a 
map satisfying the condition of Lemma \ref{le10}
for the probability distribution 
$\hat{p}_n(\omega)\defeq \frac{p_n(\omega)}{1-\epsilon_n}$
on the set $S_n(a)^c$ in the case of 
$M_n=\hat{M}_n\defeq (1-\epsilon_n) e^{na}$ and
${M'}_n= \sqrt{\hat{M}_n}$, where
$\epsilon_n \defeq p_n (S_n(a))$
and the domain of $\hat{\phi}_n$ is
$\{\tilde{M}_n +1, \ldots, \hat{M}_n+ \tilde{M}_n\}$.
Thus,
\begin{align*}
& D(p_n \circ \hat{\phi}_n^{-1}\|p_{U,\hat{{\cal M}}_n})\\
\le &
\log ((1-\epsilon_n) e^{na})
(\hat{p}_n 
\{ - \frac{1}{n} \log p_n (\omega) < a\}
+ \frac{2}{\sqrt{\hat{M}_n}}).
\end{align*}
Since any element of $S_n(a)^c$ does not satisfy
the condition $- \frac{1}{n} \log p_n (\omega) < a$,
the inequality \begin{align*}
 H(p_n \circ \hat{\phi}_n^{-1}) 
\ge na + \log (1-\epsilon_n) - 
\frac{2(na + \log (1-\epsilon_n))}{\sqrt{\hat{M}_n}}
\end{align*}
holds.

We define the code $\Phi_n=({\cal M}_n,\phi_n,\psi_n)$
with the size $M_n=  \tilde{M}_n+ \hat{M}_n$ as follows:
The encoding $\phi_n$ is defined from $\tilde{\phi}_n$ and
$\hat{\phi}_n$.
The decoding $\psi_n$ on the subset ${\cal \tilde{M}}_n$ of
${\cal M}_n$ is the inverse map of $\tilde{\phi}$.
Then, we evaluate $H(p_n \circ \phi_n^{-1})$ as
\begin{align*}
&H(p_n \circ \phi_n^{-1}) \\
=& 
H(e^{na},p_n) + (1-\epsilon_n)(  H(p_n \circ \hat{\phi}_n^{-1}) - 
\log (1-\epsilon_n)) \\
\ge &
H(e^{na},p_n) \\
&+
(1-\epsilon_n)
\Biggl(
na  -
\frac{2(na + \log (1-\epsilon_n))
}{\sqrt{\hat{M}_n}}\Biggr)\\
=&
n \int_{0}^{a}x F_n(\,d x)
+ na (1- F_n(a)) \\
&-
\frac{2(1-\epsilon_n)
(na + \log (1-\epsilon_n))}
{\sqrt{\hat{M}_n}}.
\end{align*}
Dividing both sides by $n$ and taking the limit,
we obtain the opposite inequality of (\ref{8-5-2}),
which implies the inequality of (\ref{8-5-2}).
Similar to Theorem \ref{th5},
we can prove that this code satisfies the condition
$\lim \varepsilon (\Phi_n) = \epsilon$.
\endproof

\subsection{Proof of (\ref{19-8}) in Theorem \ref{th15}}
\noindent{\em Proof of direct part:}\quad
For any real numbers $\epsilon \,> 0$ and $a$ satisfying 
\begin{align}
a \,< \underline{H} (1- e^{-\delta}|\overline{p}),\Label{19-23}
\end{align}
we construct 
a sequence $\Psi_n= ({\cal M}_n,\phi_n)$ 
such that
\begin{align*}
\vlimsup
D(p_{U,{\cal M}_n}\|p_n\circ \phi_n^{-1})
& \,< \delta \\
\vliminf \frac{1}{n}\log |\Psi_n| &= a - \epsilon.
\end{align*}
We define the probability distribution 
$\hat{p}_n(\omega)
\defeq \frac{p_n(\omega)}{p_n(S_{n}(a)^c)}
$ on 
$S_{n}(a)^c\defeq 
\{\frac{-1}{n}\log p_n (\omega) \ge a\}$
($S_{n}(a)\defeq 
\{\frac{-1}{n}\log p_n (\omega) \,< a\}$).
Since
$\hat{p}_n  \{ \hat{p}_n (\omega)\,> \frac{e^{-na}}{p_n(S_{n}(a)^c)}\}
=\hat{p}_n  \{ p_n (\omega)\,> e^{-na}\}$ and
$S_{n}(a)^c\cap \{ p_n (\omega)\,> e^{-na}\} = \emptyset$,
there exists a map from $S_{n}(a)^c$ to
$\{1, \ldots, \hat{M}_n \defeq e^{n(a-\epsilon)} p_n(S_{n}(a)^c)\}$
such that
the minimum probability of the distribution $\hat{p}_n\circ \phi_n^{-1}$
is greater than
\begin{align*}
\frac{1}{\hat{M}_n}
- \frac{e^{-na} }{p_n(S_{n}(a)^c))}
= 
\frac{1}{\hat{M}_n}
\left(
1- e^{-n\epsilon}
\right).
\end{align*}
Hence, we obtain 
\begin{align}
 D(p_{U,\hat{\cal M}_n}\|\hat{p}_n\circ \hat{\phi}_n^{-1}) 
\le & -
\log  \frac{1}{\hat{M}_n}
\left(
1- e^{-n\epsilon}
\right) + \log \frac{1}{\hat{M}_n} \nonumber \\
= &- \log \left(
1- e^{-n\epsilon}
\right)\to 0.
 \Label{19-9}
\end{align}
Next, we define a map $\phi_n$ from 
$\Omega_n$ to ${\cal M}_n= 
\{ 1, \ldots ,\hat{M}_n, \hat{M}_n+1\}$
by
$\phi_n|_{S_{n}^c(a)}= \hat{\phi}_n$ and 
$\phi_n(S_{n}(a))= \hat{M}_n+1$.
Then,
\begin{align}
&D(p_{U,{\cal M}_n}\|p_n\circ {\phi}_n^{-1}) \nonumber\\
=&
-\frac{1}{\hat{M}_n+1}
\log (\hat{M}_n+1)
+ 
\frac{\hat{M}_n}{\hat{M}_n+1}
\Biggl(
D(p_{U,\hat{\cal M}_n}\|\hat{p}_n\circ \hat{\phi}_n^{-1}) \nonumber \\
& \hspace{15ex} + \log \frac{\hat{M}_n}{\hat{M}_n+1}
- \log p_n(S_{n}(a)^c)
\Biggr). \Label{19-10}
\end{align}
Since 
\begin{align}
\vlimsup p_n(S_{n}(a))\,< 1- e^{-\delta},
\end{align}
we have the inequality (\ref{19-23}) that guarantees 
\begin{align*}
&\vlimsup D(p_{U,{\cal M}_n}\|p_n\circ {\phi}_n^{-1}) \\
=& \vlimsup -\log p_n(S_{n}(a)^c)
= \vlimsup -\log (1- p_n(S_{n}(a))) 
\,< \delta.
\end{align*}
Moreover, 
\begin{align*}
&\lim\frac{1}{n} \log |{\cal M}_n|=
\lim \frac{1}{n} \log (\hat{M}_n+1)\\
= &
\lim \frac{1}{n} \log \frac{e^{n(a-\epsilon)}}{p_n(S_{n}(a)^c)}
= a- \epsilon .
\end{align*}
\endproof

\noindent{\em Proof of converse part:}\quad
Assume that
a sequence $\Psi_n=({\cal M}_n,\phi_n)$
satisfies 
\begin{align}
\vliminf \frac{1}{n}\log |\Psi_n| &= R \Label{19-5}\\
\vlimsup
D(p_{U,{\cal M}_n}\|p_n\circ \phi_n^{-1}) &\,< \delta.\nonumber
\end{align}
For any $\epsilon'\,> 0$,
we define
\begin{align}
M_n' &\defeq | \{ 
\frac{-1}{n}\log p_n\circ \phi_n^{-1}(i)\,< R - \epsilon'
\}| \nonumber\\
\epsilon_n & \defeq
p_n \circ \phi_n^{-1}\{ 
\frac{-1}{n}\log p_n\circ \phi_n^{-1}(i)\,< R - \epsilon'
\} \nonumber\\
& \ge 
p_n \{ 
\frac{-1}{n}\log p_n(\omega)\,< R - \epsilon'
\}. \Label{19-6}
\end{align}
Information processing inequality of KL-divergence
guarantees that
\begin{align*}
& D(p_{U,{\cal M}_n}\|p_n\circ \phi_n^{-1}) \\
\ge &
\frac{M_n'}{|\Phi_n|}
\left(\log \frac{M_n'}{|\Phi_n|}- \log \epsilon_n\right) \\
& + \left(1- \frac{M_n'}{|\Phi_n|}\right)
\left(\log \left(1- \frac{M_n'}{|\Phi_n|}\right)
- \log (1-\epsilon_n)\right).
\end{align*}
Since $M_n' \le e^{n(R -\epsilon)}$ and (\ref{19-5}),
\begin{align*}
\frac{M_n'}{|\Phi_n|} \to 0, \quad 
\frac{M_n'}{|\Phi_n|}\log \frac{M_n'}{|\Phi_n|} \to 0.
\end{align*}
Therefore, taking the limit $\vlimsup$, we have
\begin{align*}
&\delta\,> \vlimsup D(p_{U,{\cal M}_n}\|p_n\circ \phi_n^{-1})
\ge \vlimsup - \log (1-\epsilon_n) \\
= &
- \log (1-\vlimsup \epsilon_n),
\end{align*}
which implies
\begin{align*}
\vlimsup \epsilon_n \,< 1- e^{-\delta}.
\end{align*}
Thus, inequality (\ref{19-6}) yields
\begin{align*}
\vlimsup p_n \{ 
\frac{-1}{n}\log p_n(\omega)\,< R - \epsilon'
\}\,< 1- e^{-\delta}.
\end{align*}
Therefore,
\begin{align*}
R - \epsilon' \le
\underline{H} (1- e^{-\delta}|\overline{p}).
\end{align*}
Since $\epsilon'$ is arbitrary,
we obtain
\begin{align*}
S^*_1(\delta|\overline{p})
\le
\underline{H} (1- e^{-\delta}|\overline{p}).
\end{align*}
\endproof

\subsection{Proof of (\ref{8-19}) in Theorem \ref{th15}}
First, by using the following two lemmas,
we will prove (\ref{8-19}).

\begin{lem}\Label{le19-2}
When three sequences of positive numbers $a_n$, $b_n$, and $c_n$
satisfy
\begin{align*}
a_n \le b_n + c_n,
\end{align*}
then
\begin{align*}
\vlimsup \frac{1}{n}\log a_n \le
\max\{ \vlimsup \frac{1}{n}\log b_n ,
\vlimsup \frac{1}{n}\log c_n \}.
\end{align*}
\end{lem}
\begin{lem}\Label{le19-1}
\begin{align}
\sup_a \{ a- \sigma(a)|\sigma(a) \,<\delta\}
\ge 
\sup_a \{ \overline{\xi}(a)|
a- \overline{\xi}(a) \,< \delta \}, \Label{8-19-1}
\end{align}
where $\overline{\xi}(a)$ is defined as:
\begin{align*}
\overline{\xi}(a)&\defeq
\vlimsup \frac{1}{n}\log |\{
\frac{-1}{n}\log p_n(\omega)
\,< a\}|.
\end{align*}
\end{lem}

\noindent{\em Proof of direct part:}\quad
We will prove 
\begin{align*}
S^*_2(\delta|\overline{p})
\ge \sup_a \{ a- \sigma(a)|\sigma(a) \,< \delta\}.
\end{align*}
That is, for any real numbers $\epsilon \,> 0$ and 
$a$ satisfying $\sigma(a) \,< \delta$,
we construct 
a sequence $\Psi_n= ({\cal M}_n,\phi_n)$ 
such that
\begin{align*}
\vlimsup
\frac{1}{n}D(p_{U,{\cal M}_n}\|p_n\circ \phi_n^{-1})
& \,< \delta \\
\vliminf \frac{1}{n}\log |\Psi_n| &= a- \sigma(a)- \epsilon.
\end{align*}
Similar to the proof of (\ref{19-8}),
we define $\hat{p}_n(\omega)$, 
$S_{n}(a)^c$, $S_{n}(a)$ and $\phi_n$.

Using (\ref{19-9}) and (\ref{19-10}),
we have
\begin{align*}
&\lim \frac{1}{n} D(p_{U,{\cal M}_n}\|p_n\circ {\phi}_n^{-1}) \\
=& \lim \frac{-1}{n}\log p_n(S_{n}(a)^c)
= \sigma(a)\,< \delta.
\end{align*}
Moreover, 
\begin{align*}
&\lim \frac{1}{n} \log |{\cal M}_n|=
\lim \frac{1}{n} \log (\hat{M}_n+1)\\
= &
\lim \frac{1}{n} \log \frac{e^{n(a-\epsilon)}}{p_n(S_{n}(a)^c)}
= a- \epsilon - \sigma(a).
\end{align*}
\endproof
\noindent{\em Proof of converse part:}\quad
We will prove 
\begin{align}
S^*_2(\delta|\overline{p})
\le 
\sup_a \{ a- \sigma(a)|\sigma(a) \,<\delta\}.
\Label{17-3}
\end{align}
That is, for any sequence $\Psi_n= ({\cal M}_n,\phi_n)$
satisfying $\vlimsup \frac{1}{n}
 D(p_{U,{\cal M}_n}\|p_n\circ {\phi}_n^{-1})\le \delta$,
we will prove that 
\begin{align*}
R\defeq \vliminf \frac{1}{n}\log |{\cal M}_n| & 
\le 
\sup_a \{ a- \sigma(a)|\sigma(a) \,<\delta\}.
\end{align*}
Let $\{n_k\}$ be a subsequence such that
$\lim_k \frac{1}{n_k}\log |{\cal M}_{n_k}| 
= \vliminf \frac{1}{n}\log |{\cal M}_n| $.
We choose the real number $a_0$
\begin{align*}
a_0 \defeq 
\inf \{ a |
\vliminf_k 
p_{U,{\cal M}_{n}}
\{\frac{-1}{n_k}\log 
p_{n_k} \circ \phi_{n_k}^{-1}(i)\,< a 
\} \,> 0\}.
\end{align*}
For any real number $\epsilon_0 \,> 0$,
the relation $\vliminf_k p_{U,{\cal M}_{n_k}}
\{\frac{-1}{n_k}\log 
p_n \circ \phi_{n_k}^{-1}(i)\,< a_0- \epsilon_0 \} = 0$
holds.
Since
\begin{align*}
& n  (a_0- \epsilon_0 )
p_{U,{\cal M}_{n_k}}
\{\frac{-1}{n_k}\log 
p_{n_k} \circ \phi_{n_k}^{-1}(i)\ge a_0- \epsilon_0 \} \\
\le & 
- \sum_i
p_{U,{\cal M}_{n_k}}(i)
\log p_{n_k}\circ \phi_{n_k}^{-1}(i),
\end{align*}
we have
\begin{align*}
& n  (a_0- \epsilon_0 )
p_{U,{\cal M}_{n_k}}
\{\frac{-1}{n_k}\log 
p_{n_k} \circ \phi_{n_k}^{-1}(i)\ge a_0- \epsilon_0 \} \\
&\qquad - \log M_{n_k}\\
\le & 
- \log M_{n_k}
- \sum_i
p_{U,{\cal M}_{n_k}}(i)
\log p_{n_k}\circ \phi_{n_k}^{-1}(i)\\
=&
D(p_{U,{\cal M}_{n_k}}
\|p_{n_k}\circ \phi_{n_k}^{-1}).
\end{align*}
Thus, 
\begin{align*}
& (a_0- \epsilon_0 )-R \\
= &
\vlimsup_k
\Biggl( (a_0- \epsilon_0 )
p_{U,{\cal M}_{n_k}}
\{\frac{-1}{n_k}\log 
p_n \circ \phi_{n_k}^{-1}(i)\ge a_0- \epsilon_0 \}\\
&\hspace{5ex} - \frac{1}{n} \log M_{n_k}\Biggr) \\
\le & \vlimsup_{k} \frac{1}{n_k} D(p_{U,{\cal M}_{n_k}}
\|p_{n_k}\circ \phi_{n_k}^{-1})
\,< \delta. 
\end{align*}
Taking the limit $\epsilon_0 \to 0$,
\begin{align*}
a_0-R \le \vlimsup_{k} \frac{1}{n_k} D(p_{U,{\cal M}_{n_k}}
\|p_{n_k}\circ \phi_{n_k}^{-1})
\,< \delta.
\end{align*}

Next, we choose a real number $\epsilon$ such that
\begin{align}
0 \,< \epsilon \,< \delta - (a_0 - R).
\Label{17-1}
\end{align}
Then, there exits a real number $\alpha \,>0$
such that
\begin{align*}
\vliminf
p_{U,{\cal M}_{n_k}}
\{\frac{-1}{n}\log 
p_{n_k} \circ \phi_{n_k}^{-1}(i)\,< a_0 + \epsilon 
\} 
\,> \alpha.
\end{align*}
Thus,
\begin{align*}
| \{\frac{-1}{n_k}\log 
p_{n_k} \circ \phi_{n_k}^{-1}(i)\,< a_0 + \epsilon 
\}|
\,> \alpha M_{n_k}
\end{align*}
for sufficiently large $n_k$.
Since
\begin{align*}
& p_n
\Biggl(\{\frac{-1}{n_k}\log 
p_{n_k} \circ \phi_{n_k}^{-1}(i)\,< a_0 + \epsilon 
\}\\
& \hspace{5ex} \setminus
\phi_{n_k} \{\frac{-1}{n_k}\log 
p_{n_k} (\omega )\,< a_0 + \epsilon 
\}
\Biggr)\\
\le &
p_n\{\frac{-1}{n_k}\log 
p_{n_k} (\omega )\ge a_0 + \epsilon \},
\end{align*}
we can evaluate
\begin{align*}
& \Biggl|\{\frac{-1}{n_k}\log 
p_{n_k} \circ \phi_{n_k}^{-1}(i)\,< a_0 + \epsilon 
\}\\
& \hspace{5ex}\setminus
\phi_{n_k} \{\frac{-1}{n_k}\log 
p_{n_k} (\omega )\,< a_0 + \epsilon 
\}\Biggr|\\
\le &
\frac{p_n\{\frac{-1}{n_k}\log 
p_{n_k} (\omega )\ge a_0 + \epsilon \}}
{e^{-n(a_0 + \epsilon)}}.
\end{align*}
Thus,
\begin{align*}
& \alpha M_{n_k} \,< |\{\frac{-1}{n_k}\log 
p_{n_k} \circ \phi_{n_k}^{-1}(i)\,< a_0 + \epsilon \}| \\
\le &
\frac{p_n\{\frac{-1}{n_k}\log 
p_{n_k} (\omega )\ge a_0 + \epsilon \}}
{e^{-n(a_0 + \epsilon)}}+
|\{\frac{-1}{n_k}\log 
p_{n_k} (\omega )\,< a_0 + \epsilon \}|.
\end{align*}
Using Lemma \ref{le19-2},
we have
\begin{align}
\max\{\overline{\xi}(a_0+\epsilon),
(a_0+\epsilon)- \sigma(a_0+\epsilon)\}
\ge R \Label{17-2}.
\end{align}

If
$\overline{\xi}(a_0+\epsilon) \ge (a_0+\epsilon)- \sigma(a_0+\epsilon)$,
by combining (\ref{17-1}) and (\ref{17-2}),
we can show
\begin{align*}
(a+ \epsilon)- \overline{\xi}(a_0+\epsilon)\,< \delta.
\end{align*}
Therefore, 
we obtain
\begin{align*}
R \le 
\sup_a \{ \overline{\xi}(a)|
a- \overline{\xi}(a) \,< \delta \}
\le 
\sup_a \{ a- \sigma(a)|\sigma(a) \,<\delta\}.
\end{align*}

If
$\overline{\xi}(a_0+\epsilon) \,< (a_0+\epsilon)- \sigma(a_0+\epsilon)$,
combining (\ref{17-1}) and (\ref{17-2}),
we can show
\begin{align*}
\sigma(a_0+ \epsilon)\,< \delta.
\end{align*}
Therefore,
we obtain
\begin{align*}
R \le \sup_a 
\{ a- \sigma(a)|\sigma(a) \,<\delta\}.
\end{align*}
\endproof
\noindent{\em Proof of Lemma \ref{le19-2}:}\quad
Since
\begin{align*}
b_n+c_n \le \max\{ 2 b_n , 2 c_n\},
\end{align*}
we have
\begin{align*}
\frac{1}{n}\log a_n
\le \max\{
\frac{\log 2}{n}+ \frac{1}{n}\log b_n, 
\frac{\log 2}{n}+ \frac{1}{n}\log c_n\}.
\end{align*}
Taking the limit $\vlimsup$,
we obtain
\begin{align*}
\vlimsup \frac{1}{n}\log a_n
\le \max\{
\vlimsup \frac{1}{n}\log b_n, 
\vlimsup \frac{1}{n}\log c_n \}.
\end{align*}
\endproof
\noindent{\em Proof of Lemma \ref{le19-1}:}\quad
In this proof, the following lemma plays an important role.
\begin{lem}\Label{le17}Hayashi\cite[Lemma 13]{ha}
If two decreasing functions $f$ and $g$ satisfy
\begin{align}
-f(a) + a \ge g (b) \hbox{ if }
f(a) \,> f(b),\Label{c3}
\end{align}
then
\begin{align*}
 \sup_a \{a- g(a)|g(a) \,< \delta\}
\ge 
\sup_a \{ f(a)|
a- f(a) \,< \delta \}.
\end{align*}
\end{lem}
\begin{rem}
This lemma is essentially the one obtained by Hayashi\cite{ha}.
But, this statement is a little different from Hayashi\cite{ha}'s.
\end{rem}
\begin{proof}
We prove Lemma \ref{le17} by reduction to absurdity.
Assume that there exists a real number $a_0$ such that
\begin{align}
a_0 -f (a_0) & \,< \delta , \Label{la1}\\
f (a_0) & \,>
\sup_a\{a- g(a) | g(a)\le r \}.\Label{la2}
\end{align}
We define 
$a_1 := \inf_a \{ a |f (a)=f(a_0)\}$ and
assume that 
$a_0 \,> a_1$.
For any real number $\epsilon: 0 \,< \epsilon \,< a_0 - a_1$,
the inequality $f(a_1- \epsilon)
\,< f(a_1 + \epsilon)
$ holds. Using (\ref{c3}), we have
\begin{align*}
& g(a_1- \epsilon)
\le
-f(a_1+ \epsilon) + a_1 + \epsilon
=
-f(a_0) + a_1 +\epsilon \\
\,< &
\delta +(a_1-a_0) + \epsilon 
\,< \delta
\end{align*}
Thus,
\begin{align*}
& \sup_a\{a-g(a) | g(a)\,< \delta \}
\ge 
a_1-\epsilon - g(a_1- \epsilon) \\
\ge &
a_1-\epsilon - (a_1+\epsilon)+ f(a_1+ \epsilon) 
=  f(a_0)- 2\epsilon.
\end{align*}
Taking the limit $\epsilon  \to 0$,
we obtain
$\sup\{a- g (a) | g (a)\,< r \}\ge f(a_0) $,
which contradicts (\ref{la2}).

Next, we treat the case $a_0 =a_1$.
The inequality $f (a_0) \,>
f (a_0 - \epsilon)$
holds for $\forall \epsilon \,> 0$.
Using (\ref{c3}),
we have $g(a_0 - \epsilon) \le 
-f (a_0) + a_0 \le \delta$.
Thus,
\begin{align*}
& \sup_a\{a- g(a) | g(a)\le r \}
\ge
a_0 - \epsilon - g(a_0 - \epsilon) \\
\ge & a_0 - \epsilon - a_0 + f(a_0)
= -\epsilon + f(a_0).
\end{align*}
This also contradicts (\ref{la2}).
\end{proof}
Since 
\begin{align*}
(p_n- e^{na})\{ p_n- e^{na} \le 0\}
\le 
(p_n- e^{na})\{ p_n- e^{nb} \le 0\}.
\end{align*}
By adding $e^{na}$ to both sides,
we have
\begin{align*}
& p_n \{ p_n- e^{na} \le 0\}
+ e^{na}|\{ p_n- e^{na} \,> 0\}|\\
\le &
p_n \{ p_n- e^{nb} \le 0\}
+ e^{na}|\{ p_n- e^{nb} \,> 0\}|,
\end{align*}
which implies
\begin{align*}
& |\{ p_n- e^{na} \,> 0\}| \\
 \le &
e^{-na} p_n \{ p_n- e^{nb} \le 0\}
+ |\{ p_n- e^{nb} \,> 0\}| .
\end{align*}
Thus, Lemma \ref{le19-2} guarantees that
\begin{align*}
\overline{\xi}(a)\le \max\{ - a - \sigma(b),
\overline{\xi}(b)\}.
\end{align*}
Using this relation, we obtain
\begin{align*}
\overline{\xi}(a)\le - a - \sigma(b)\hbox{ if }
\overline{\xi}(b) \,< \overline{\xi}(a).
\end{align*}
Therefore, by applying Lemma \ref{le17} to
the case of $f= \overline{\xi}, g=\sigma$,
we can show (\ref{8-19-1}).
\endproof

\subsection{Proof of Theorem \ref{th91}}
\noindent{\em Proof of inequality (\ref{8-5-2}):}\quad
We define the probability distribution function
$F_n$ on the real numbers
$\real$ as:
\begin{align}
F_n(x)\defeq p_n\{ -\frac{1}{n} \log p_n (\omega) \,< 
\overline{H}(\overline{p})+ \frac{x}{\sqrt{n}} \}
\end{align}
for a probability distribution $p_n$.
Then,
the relation 
\begin{align}
H(M_n,p_n)=
\int_{0}^{b_n}
(\sqrt{n} x + n \overline{H}(\overline{p}))
F_n(\,d x)\Label{8-13-2}
\end{align}
holds,
where $b_n \defeq \frac{1}{\sqrt{n}}
(\log M_n - n \overline{H}(\overline{p}))$.
Thus, Lemma \ref{8-6-1} yields the inequality
\begin{align*}
& H(p_n\circ \phi_n^{-1})\\
\le & \int_{0}^{b_n}
(\sqrt{n} x + n \overline{H}(\overline{p}))
F_n(\,d x)\\
&+
p_n\{ p_n (\omega) \le \frac{1}{M_n} \} \\
& \quad \times (\sqrt{n} b_n + n \overline{H}(\overline{p})- 
\log p_n\{ p_n (\omega) \le \frac{1}{M_n} \}) \\
= & \sqrt{n} 
\int_{0}^{b_n}
x F_n(\,d x)+ n \overline{H}(\overline{p})\\
&+
p_n\{ p_n (\omega) \le \frac{1}{M_n} \}
(\sqrt{n} b_n - \log p_n\{ p_n (\omega) \le \frac{1}{M_n} \}) .
\end{align*}
Therefore,
the inequality
\begin{align*}
& \frac{1}{\sqrt{n}}
\left(
H(p_n\circ \phi_n^{-1})
- n \overline{H}(\overline{p})
\right) \\
\le &
\int_{0}^{b_n}
x F_n(\,d x)\\
&+
p_n\{ p_n (\omega) \le \frac{1}{M_n} \}
(\sqrt{n} b_n - \log p_n\{ p_n (\omega) \le \frac{1}{M_n} \}) 
\end{align*}
holds.
Taking the limit $\vliminf$, we obtain 
(\ref{8-5-5-1}),
which is equivalent with (\ref{8-5-2}).

\noindent{\em Proof of the existence part:}\quad
In the following, by using Lemma \ref{le10}, 
we construct the code $\Phi_n= ({\cal M}_n,\phi_n,\psi_n)$
satisfying the equality at (\ref{8-5-5-1}) and
$\vliminf \varepsilon (\Phi_n)= \epsilon$ as follows.
Let $\tilde{\phi}_n$ be the one-to-one map
from 
\begin{align*}
S_n(\overline{H}(\overline{p}),b)\defeq
\{ - \frac{1}{n} \log p_n (\omega) < 
\overline{H}(\overline{p}) +\frac{b}{\sqrt{n}} \}
\end{align*}
to ${\cal \tilde{M}}_n\defeq
\{1, \ldots, \tilde{M}_n\}$,
where $\tilde{M}_n\defeq |S_n(\overline{H}(\overline{p}),b) |$.
Then, the inequality 
$\tilde{M}_n \le e^{n\overline{H}(\overline{p}) +b\sqrt{n}}$ 
holds.
Furthermore, 
we define $\hat{\phi}_n$ as a 
map satisfying the condition of Lemma \ref{le10}
for the probability distribution 
$\hat{p}_n(\omega)\defeq \frac{p_n(\omega)}{1-\epsilon_n}$
on the set $S_n(\overline{H}(\overline{p}),b)^c$ in the case of 
$M_n=\hat{M}_n\defeq (1-\epsilon_n) 
e^{n\overline{H}(\overline{p}) +b\sqrt{n}}$ and
${M'}_n= \sqrt{\hat{M}_n}$, where
$\epsilon_n \defeq p_n (S_n(\overline{H}(\overline{p}),b))$
and the domain of $\hat{\phi}_n$ is
$\{\tilde{M}_n +1, \ldots, \hat{M}_n+ \tilde{M}_n\}$.
Thus,
\begin{align*}
& D(p_n \circ \hat{\phi}_n^{-1}\|p_{U,\hat{{\cal M}}_n})\\
\le &
\log ((1-\epsilon_n) e^{n\overline{H}(\overline{p}) +b\sqrt{n}})
\cdot \\
& (\hat{p}_n 
\{ - \frac{1}{n} \log p_n (\omega) < 
\overline{H}(\overline{p}) +\frac{b}{\sqrt{n}}\}
+ \frac{2}{\sqrt{\hat{M}_n}}).
\end{align*}
Because no element of $S_n(\overline{H}(\overline{p}),b)^c$ 
satisfies the condition
$- \frac{1}{n} \log p_n (\omega) < 
\overline{H}(\overline{p}) +\frac{b}{\sqrt{n}}$,
the inequality
\begin{align*}
 H(p_n \circ \hat{\phi}_n^{-1}) 
\ge & \log (1- \epsilon_n)
+ n \overline{H}(\overline{p}) +  \sqrt{n}b \\
&-(n \overline{H}(\overline{p}) +  \sqrt{n}b + \log (1-\epsilon_n))
\frac{2}{\sqrt{\hat{M}_n}}
\end{align*}
holds.

We define the code $\Phi_n=({\cal M}_n,\phi_n,\psi_n)$
with the size $M_n=  \tilde{M}_n+ \hat{M}_n$ similar to
the proof of Theorem \ref{th9}.
Then,
\begin{align*}
&H(p_n \circ \phi_n^{-1}) \\
= & 
H(e^{n\overline{H}(\overline{p}) +b\sqrt{n}},p_n) 
+ (1-\epsilon_n)(  H(p_n \circ \hat{\phi}_n^{-1}) - 
\log (1-\epsilon_n)) \\
\ge &
H(e^{n\overline{H}(\overline{p}) +b\sqrt{n}},p_n) 
+
(1-\epsilon_n)
\Biggl(
 n \overline{H}(\overline{p}) +  \sqrt{n}b \\
&-(n \overline{H}(\overline{p}) +  \sqrt{n}b + \log (1-\epsilon_n))
\frac{2}{\sqrt{\hat{M}_n}}\Biggr)\\
= &
\sqrt{n}\int_{0}^{b}
x F_n(\,d x)
+ n \overline{H}(\overline{p})
+ \sqrt{n} b(1-F_n(b)) \\
&-\frac{2(1-\epsilon_n)(n \overline{H}(\overline{p}) +  \sqrt{n}b + \log (1-\epsilon_n))}
{\sqrt{\hat{M}_n}}.
\end{align*}
By substracting $n \overline{H}(\overline{p})$ from both sides,
dividing both by $\sqrt{n}$, and
taking the limit,
we obtain the opposite inequality of (\ref{8-5-5-1}),
which implies the inequality of (\ref{8-5-5-1}).
Similar to Theorem \ref{th5},
we can prove that this code satisfies the condition
$\lim \varepsilon (\Phi_n) = \epsilon$.
\endproof

\subsection{Proof of Theorem \ref{th20}}
\noindent{\em Proof of direct part:}\quad
For for any real numbers $\epsilon \,> 0$ and $a$ satisfying 
\begin{align}
b \,< \underline{H} (1- e^{-\delta},a|\overline{p}),\Label{19-22}
\end{align}
we construct 
a sequence $\Psi_n= ({\cal M}_n,\phi_n)$ 
such that
\begin{align*}
\vlimsup D(p_{U,{\cal M}_n}\|p_n\circ \phi_n^{-1})
& \,< \delta \\
\vliminf \frac{1}{\sqrt{n}}\log \frac{|\Psi_n|}{e^{na}} &= b - \epsilon.
\end{align*}
We define the probability distribution 
$\hat{p}_n(\omega)
\defeq \frac{p_n(\omega)}{p_n(S_{n}(a,b)^c)}
$ on 
$S_{n}(a,b)^c\defeq 
\{\frac{-1}{n}\log p_n (\omega) \ge a +\frac{b}{\sqrt{n}}\}$
($S_{n}(a,b)\defeq 
\{\frac{-1}{n}\log p_n (\omega) \,< a+\frac{b}{\sqrt{n}}\}$).
Then, for any $\epsilon\,> 0$, 
similar to our proof of (\ref{19-8}) in Theorem \ref{th15},
there exists an operation $\hat{\phi}_n$ from $S_{n}(a,b)^c$ to 
$\hat{\cal M}_n\defeq e^{na +\sqrt{n}(b-\epsilon)}p_n(S_{n}(a,b)^c)$
such that 
\begin{align*}
D(p_{U,\hat{\cal M}_n}\|\hat{p}_n\circ \hat{\phi}_n^{-1}) 
\le - \log (1- e^{-\epsilon \sqrt{n}} )
\to 0. 
\end{align*}
Next, we define a map $\phi_n$ from 
$\Omega_n$ to ${\cal M}_n= 
\{ 1, \ldots ,\hat{M}_n, \hat{M}_n+1\}$
by
$\phi_n|_{S_{n}^c(a,b)}= \hat{\phi}_n$ and 
$\phi_n(S_{n}(a,b))= \hat{M}_n+1$.
Then, we obtain
\begin{align}
&D(p_{U,{\cal M}_n}\|p_n\circ {\phi}_n^{-1}) \nonumber\\
=&
-\frac{1}{\hat{M}_n+1}
\log (\hat{M}_n+1)
+ 
\frac{\hat{M}_n}{\hat{M}_n+1}
\Biggl(
D(p_{U,\hat{\cal M}_n}\|\hat{p}_n\circ \hat{\phi}_n^{-1}) \nonumber \\
& \hspace{15ex} + \log \frac{\hat{M}_n}{\hat{M}_n+1}
- \log p_n(S_{n}(a,b)^c)
\Biggr). \nonumber 
\end{align}
Since the inequality (\ref{19-22}) guarantees 
\begin{align}
\vlimsup p_n(S_{n}(a,b))\,< 1- e^{-\delta},
\end{align}
we have
\begin{align*}
&\vlimsup D(p_{U,{\cal M}_n}\|p_n\circ {\phi}_n^{-1}) \\
=& \vlimsup -\log p_n(S_{n}(a,b)^c)
= \vlimsup -\log (1- p_n(S_{n}(a,b))) 
\,< \delta.
\end{align*}
Moreover, 
\begin{align*}
&\lim\frac{1}{\sqrt{n}} \log |{\cal M}_n|=
\lim \frac{1}{\sqrt{n}} \log (\hat{M}_n+1)\\
= &
\lim \frac{1}{\sqrt{n}} \log \frac{e^{n(a-\epsilon)}}{p_n(S_{n}(a,b)^c)}
= b- \epsilon .
\end{align*}
\endproof

\noindent{\em Proof of converse part:}\quad
Assume that
a sequence $\Psi_n=({\cal M}_n,\phi_n)$
satisfies 
\begin{align}
\vliminf \frac{1}{\sqrt{n}}\log \frac{|\Psi_n|}{e^{na}} &= R \Label{19-5-1}\\
\vlimsup
D(p_{U,{\cal M}_n}\|p_n\circ \phi_n^{-1}) &\,< \delta.\nonumber
\end{align}
For any $\epsilon'\,> 0$,
we define
\begin{align}
M_n' &\defeq | \{ 
\frac{-1}{n}\log p_n\circ \phi_n^{-1}(i)\,< a + \frac{R - \epsilon'}{\sqrt{n}}
\}| \nonumber\\
\epsilon_n & \defeq
p_n \circ \phi_n^{-1}\{ 
\frac{-1}{n}\log p_n\circ \phi_n^{-1}(i)\,< a + \frac{R - \epsilon'}{\sqrt{n}}
\} \nonumber\\
& \ge 
p_n \{ 
\frac{-1}{n}\log p_n(\omega)\,< a + \frac{R - \epsilon'}{\sqrt{n}}
\}. \Label{19-6-1}
\end{align}
Information processing inequality of KL-divergence
guarantees that
\begin{align*}
& D(p_{U,{\cal M}_n}\|p_n\circ \phi_n^{-1}) \\
\ge &
\frac{M_n'}{|\Phi_n|}
\left(\log \frac{M_n'}{|\Phi_n|}- \log \epsilon_n\right) \\
& + \left(1- \frac{M_n'}{|\Phi_n|}\right)
\left(\log \left(1- \frac{M_n'}{|\Phi_n|}\right)
- \log (1-\epsilon_n)\right).
\end{align*}
Since $M_n' \le e^{na +\sqrt{n}(R -\epsilon)}$ and (\ref{19-5-1}),
\begin{align*}
\frac{M_n'}{|\Phi_n|} \to 0, \quad 
\frac{M_n'}{|\Phi_n|}\log \frac{M_n'}{|\Phi_n|} \to 0.
\end{align*}
Therefore, taking the limit $\vlimsup$, we have
\begin{align*}
&\delta\,> \vlimsup D(p_{U,{\cal M}_n}\|p_n\circ \phi_n^{-1})
\ge \vlimsup - \log (1-\epsilon_n) \\
= &
- \log (1-\vlimsup \epsilon_n),
\end{align*}
which implies
\begin{align*}
\vlimsup \epsilon_n \,< 1- e^{-\delta}.
\end{align*}
Thus, the inequality (\ref{19-6-1}) yields
\begin{align*}
\vlimsup p_n \{ 
\frac{-1}{n}\log p_n(\omega)\,< R - \epsilon'
\}\,< 1- e^{-\delta}.
\end{align*}
Therefore,
\begin{align*}
R - \epsilon' \le
\underline{H} (1- e^{-\delta},a|\overline{p}).
\end{align*}
Since $\epsilon'$ is arbitrary,
we obtain
\begin{align*}
S^*_1(\delta,a|\overline{p})
\le
\underline{H} (1- e^{-\delta},a|\overline{p}).
\end{align*}
\endproof

\subsection{Proof of Theorem \ref{21-8}}\Label{22-1}
This theorem is proved by the type method.
Let ${\cal T}_n$ be the set of $n$-th types, {\em i.e.},
the set of empirical distributions of $n$ observations.
We denote the set of elements $\Omega^n$ corresponding to $P$ by
$T_P^n\subset \Omega^n$,
and define a subset $T_n(a,b)$ of the set $\Omega^n$ as
\begin{align*}
T_n(a,b)\defeq \cup_{P \in {\cal T}_n: |T_P^n|\le e^{an +b\sqrt{n}}}
T_P^n.
\end{align*}
Using this notation, we define
the encoding $\psi_n$ from $\Omega^n$ to $T_n(a,b) \cup \{0\}$:
\begin{align*}
\omega \mapsto \left\{
\begin{array}{cl}
\omega & \hbox{ if }\omega \in T_n(a,b)\\
0 & \hbox{ if }\omega \notin T_n(a,b)
\end{array}
\right. .
\end{align*}
We also define the decoding $\psi_n$ such that
$\psi_n(\omega)= \omega , \forall \omega \in T_n(a,b)$.
The relation 
\begin{align*}
\varepsilon_{P^n}(\Phi_n)= 
1- P^n (T_n(a,b))
\end{align*}
holds.
Then, the type counting lemma guarantees that
\begin{align*}
|T_n(a,b)| \le (n+1)^d e^{an +b\sqrt{n}},
\end{align*}
which implies 
\begin{align}
\liminf \frac{1}{\sqrt{n}}\log \frac{|T_n(a,b)|}{e^{na}} \le b \Label{21-1}.
\end{align}
On the other hand, 
the set $\{ - \log P^n(\omega) \,< n a + b \sqrt{n}\}$ 
can be expressed as
\begin{align*}
&\{ - \log P^n(\omega) \,< n a + b \sqrt{n}\}
= \{ P^n(\omega) \,>  e^{-n a - b \sqrt{n}}\}\\
=& 
\bigcup_{P' \in {\cal T}_n: P^n (\omega) \,>  e^{-n a - b \sqrt{n}}
\hbox{ for }\omega \in T_{P'}^n }
T_P^n.
\end{align*}
Hence, 
when a type $P' \in {\cal T}_n$ 
satisfies 
$P^n (\omega) \,>  e^{-n a - b \sqrt{n}}$ for $\omega \in T_{P'}^n $,
the inequality
$P^n(T_{P'}^n)\le 1$ yields
\begin{align*}
|T_{P'}^n| \le P^n (\omega) ^{-1} \le 
e^{n a + b \sqrt{n}}.
\end{align*}
Thus,
\begin{align*}
\{ - \log P^n(\omega) \,< n a + b \sqrt{n}\}
\subset T_n(a,b).
\end{align*}
Therefore, if the probability distribution $P$ satisfies
$H(P)=a$, then
\begin{align*}
&\Phi(\frac{b}{\sqrt{V_P}})
= \lim P^n \{ - \log P^n(\omega) \,< n a + b \sqrt{n}\}\\
\le &
\lim P^n(T_n(a,b))
=1- \lim\varepsilon_{P^n}(\Phi_n),
\end{align*}
i.e.,
\begin{align}
\lim\varepsilon_{P^n}(\Phi_n)\le 1- \Phi(\frac{b}{\sqrt{V_P}})\Label{21-2}.
\end{align}
Since the r.h.s. of (\ref{21-2}) is optimal under the condition (\ref{21-1}),
the inequality of (\ref{21-2}) holds.
Conversely, Since the r.h.s. of (\ref{21-1}) 
is optimal under the condition (\ref{21-2}),
the inequality of (\ref{21-1}) holds.
Thus, we obtain (\ref{20-3}).
\endproof

In the universal variable-length source code,
the order of the second term regarding expected coding length
is $\log n$.
But, as discussed in the above proof,
this term is negligible concerning the second order asymptotics
of fixed-length source coding.

Thus, in the variable-length and fixed-length source coding,
the central limit theorem plays an important role,
while its applications to the respective problems are
different.

\subsection{Proof of Theorem \ref{21-9}}
Using the type method, we define
a map $\phi_n$ from
$\Omega^n$ to
${\cal M}_n \defeq \{1, \ldots, \frac{1}{n} e^{n a + b \sqrt{n}}\}$
as follows.
The map $\phi_n$ maps any element of $T_n(a,b)$ to $1$.
On the other hand,
the map $\psi$ restricted to a subset $T_{P'}^n \subset T_n(a,b)^c$
is defined as
the map from $T_{P'}^n$ to ${\cal M}_n$
satisfying the conditions Lemma \ref{le1}
in the case of $M_n'= |T_{P'}^n|$.

Then, the equality of (\ref{21-2})
guarantees 
\begin{align*}
& \varepsilon_{P^n}(\Psi_n) \\
\le &\sum_{T_{P'}^n\subset T_n(a,b)^c}
P^n(T_{P'}^n)\frac{e^{n a + b \sqrt{n}}}{n |T_{P'}^n|}
+ 
\sum_{T_{P'}^n\subset T_n(a,b)}P^n(T_{P'}^n)\\
\le &
P^n(T_n(a,b)^c)\frac{1}{n}
+ P^n(T_n(a,b)) \\
\to &
\left\{ 
\begin{array}{cl}
0 & H(P) \,> a \\
\Phi(\frac{b}{\sqrt{V_P}}) & H(P) = a.
\end{array}
\right.
\end{align*}
Therefore, we obtain (\ref{20-5}).
\endproof

\section{Concluding remarks and Future study}
We proved that Folklore for source coding does not hold
for the variational distance criterion (\ref{20-1})
nor the KL-divergence criterion (\ref{19-26}) nor (\ref{19-27}).
Of course, since
our criteria (\ref{20-1}), (\ref{19-26}) and (\ref{19-27})
are more restrictive than Han's criterion (\ref{eq68}),
there is no contradiction.
But, it is necessary to discuss
which criterion is more suitable for treating
Folklore for source coding.
This is left to future research.

While we focused on the relation between 
source coding and intrinsic randomness only in the fixed-length case,
the compression scheme used in practice is variable-length.
In the variable-length setting,
if we use 
the code whose coding length is decided only from the
empirical distribution (this code is called Lynch-Davisson code)  
in the i.i.d.\ case,
the conditional distribution of the obtained data is 
the uniform distribution.
That is, in the variable-length setting,
there exists a code attaining the entropy rate with no error 
in both settings.
Thus, 
a result different from the fixed-length setting can be expected in 
the the variable-length setting.

Furthermore, this type second order asymptotics can be
extended to other topics in information theory.
Indeed, in the case of 
channel coding, resolvability, and simple hypothesis testing,
lemmas corresponding to Lemmas \ref{le3}--\ref{le2}
have been obtained by Han \cite{Han1}.
Thus, it is not difficult to derive theorems corresponding to 
Theorem \ref{th4}.
However, in channel coding 
it is difficult to calculate
the quantities corresponding to
$\underline{H}(\epsilon,a|\overline{p})$ and 
$\overline{H}(\epsilon,a|\overline{p})$ even in the i.i.d.\ case.
On the other hand,
similar to fixed-length source coding and 
intrinsic randomness, 
we can treat the second order asymptotics concerning the other two problems
in the i.i.d.\ case.
Especially, when we discuss simple hypothesis testing 
with hypothesis $p$ and $q$ from 
the second order asymptotics viewpoint,
we optimize the second order coefficient $b$ of the first error
$e^{-n D(p\|q)- \sqrt{n}b}$
under the constraint that 
the second error probability is less than the fixed constant $\epsilon$.
There is no difficulty in this problem.
However, there is considerable difficulty 
in the quantum setting of this problem.

In addition, third order asymptotics is expected, 
but it seems difficult.
In this extension of the i.i.d.\ case,
our issue is 
the difference of $\sqrt{n}(-\frac{1}{n}\log P^n- H(P))$ from 
the normal distribution.
If the next order is a constant term of $\log P^n$,
we cannot use methods similar to those described in this paper.
This is an interesting future problem.

\section*{Acknowledgments}
The author would like to thank Professor Hiroshi Imai of the QCI project for support.
He is grateful to Mr. Tsuyoshi Ito and
Dr. Mitsuru Hamada for 
useful discussions.
He also appreciates reviewers' helpful comments.

\appendix

\noindent{\em Proof of (\ref{8-6-30}) $\Rightarrow$ (\ref{8-6-31}):}\quad
The relations 
\begin{align*}
&D(p_n\circ \phi_n^{-1}\| p_{U,\cM_n})
= \log M_n - H(p_n\circ \phi_n^{-1}) \\
=& H(p_{U,\cM_n})-H(p_n\circ \phi_n^{-1})
\end{align*}
hold.

If $d(p_n\circ \phi_n^{-1}, p_{U,\cM_n}) \le 1/4$,
Fannes' inequality \cite{Fa}
(See also Csisz\'{a}r and K\"{o}rner \cite{CK})
implies 
\begin{align*}
&|H(p_{U,\cM_n})-H(p_n\circ \phi_n^{-1})| \\
\le & - d(p_n\circ \phi_n^{-1}, p_{U,\cM_n})
\log (d(p_n\circ \phi_n^{-1}, p_{U,\cM_n})/M_n).
\end{align*}

Dividing the above by $n$,
we have
\begin{align*}
& \frac{1}{n}D(p_n\circ \phi_n^{-1}\| p_{U,\cM_n}) \\
\le & 
d(p_n\circ \phi_n^{-1}, p_{U,\cM_n})
\frac{1}{n}(\log M_n 
-\log (d(p_n\circ \phi_n^{-1}, p_{U,\cM_n})).
\end{align*}
Since $\vlimsup \frac{1}{n}\log M_n < \infty$,
we obtain (\ref{8-6-30}) $\Rightarrow$ (\ref{8-6-31}).

\bibliographystyle{IEEE}

\end{document}